\documentclass[final,3p,times]{elsarticle}
\biboptions{sort&compress}
\usepackage{wrapfig}
\usepackage[numbers]{natbib}
\usepackage{natbib}

\usepackage{graphics}
\usepackage{graphicx}
\usepackage{epsfig}
\usepackage{amssymb}
\usepackage{amsthm}
\usepackage{amsmath}

\usepackage{caption}
\usepackage[labelfont=bf]{caption}
\usepackage[labelsep=space]{caption}
\usepackage[nodots]{numcompress}

\usepackage{xcolor}
\usepackage{hyperref}
\hypersetup{
  colorlinks,
  citecolor=[violet],
  linkcolor=[red],
  urlcolor=[blue]}

\journal{}

\begin{document}

\begin{frontmatter}

\title{Crystal growth engineering and origin of the weak ferromagnetism in antiferromagnetic matrix of orthochromates from $t$-$e$ orbital hybridization}

\author[1,2,3,4]{Yinghao Zhu\fnref{fn1}}
\author[1,4]{Junchao Xia\fnref{fn1}}
\author[1,4]{Si Wu\fnref{fn1}}
\fntext[fn1]{These authors contributed equally.}
\author[1]{Kaitong Sun}
\author[2]{Yuewen Yang}
\author[2]{Yanling Zhao}
\author[2]{Hei Wun Kan}
\author[3]{Yang Zhang}
\author[3]{Ling Wang}
\author[3]{Hui Wang}
\author[3]{Jinghong Fang}
\author[3]{Chaoyue Wang}
\author[3]{Tong Wu}
\author[3]{Yun Shi}
\author[3]{Jianding Yu\corref{cor1}}
\ead{yujianding@mail.sic.ac.cn}
\author[2]{Ruiqin Zhang\corref{cor1}}
\ead{aprqz@cityu.edu.hk}
\author[1]{Hai-Feng Li\corref{cor1}\fnref{fn2}}
\cortext[cor1]{Corresponding author}
\ead{haifengli@um.edu.mo}
\fntext[fn2]{Lead Author.}
\affiliation[1]{organization={Joint Key Laboratory of the Ministry of Education, Institute of Applied Physics and Materials Engineering, University of Macau},
            addressline={Avenida da Universidade, Taipa},
            city={Macao SAR},
            postcode={999078},
            country={China}}
\affiliation[2]{organization={Department of Physics, City University of Hong Kong},
            addressline={Kowloon},
            city={Hong Kong SAR},
            postcode={999077},
            country={China}}
\affiliation[3]{organization={State Key Laboratory of High Performance Ceramics and Superfine Microstructure, Shanghai Institute of Ceramics, Chinese Academy of Sciences},
            city={Shanghai},
            postcode={200050},
            country={China}}
\affiliation[4]{organization={Guangdong--Hong Kong--Macao Joint Laboratory for Neutron Scattering Science and Technology},
            addressline={No. 1. Zhongziyuan Road, Dalang},
            city={DongGuan},
            postcode={523803},
            country={China}}

\begin{abstract}
We report a combined experimental and theoretical study on intriguing magnetic properties of quasiferroelectric orthochromates. Large single crystals of the family of RECrO$_3$ (RE = Y, Eu, Gd, Tb, Dy, Ho, Er, Tm, Yb, and Lu) compounds were successfully grown. Neutron Laue study indicates a good quality of the obtained single crystals. Applied magnetic-field and temperature dependent magnetization measurements reveal their intrinsic magnetic properties, especially the antiferromagnetic (AFM) transition temperatures. Density functional theory studies of the electronic structures were carried out using the Perdew-Burke-Ernzerhof functional plus Hubbard $U$ method. Crystallographic information and magnetism were theoretically optimized systematically. When RE$^{3+}$ cations vary from Y$^{3+}$ and Eu$^{3+}$ to Lu$^{3+}$ ions, the calculated $t$-$e$ orbital hybridization degree and N\'{e}el temperature behave similarly to the experimentally-determined AFM transition temperature with variation in cationic radius. We found that the $t$-$e$ hybridization is anisotropic, causing a magnetic anisotropy of Cr$^{3+}$ sublattices. This was evaluated with the nearest-neighbour $J_1$-$J_2$ model. Our research provides a picture of the electronic structures during the $t$-$e$ hybridization process while changing RE ions and sheds light on the nature of the weak ferromagnetism coexisting with predominated antiferromagnetism. The available large RECrO$_3$ single crystals build a platform for further studies of orthochromates.
\end{abstract}

\begin{graphicalabstract}
\includegraphics[width=0.88\textwidth]{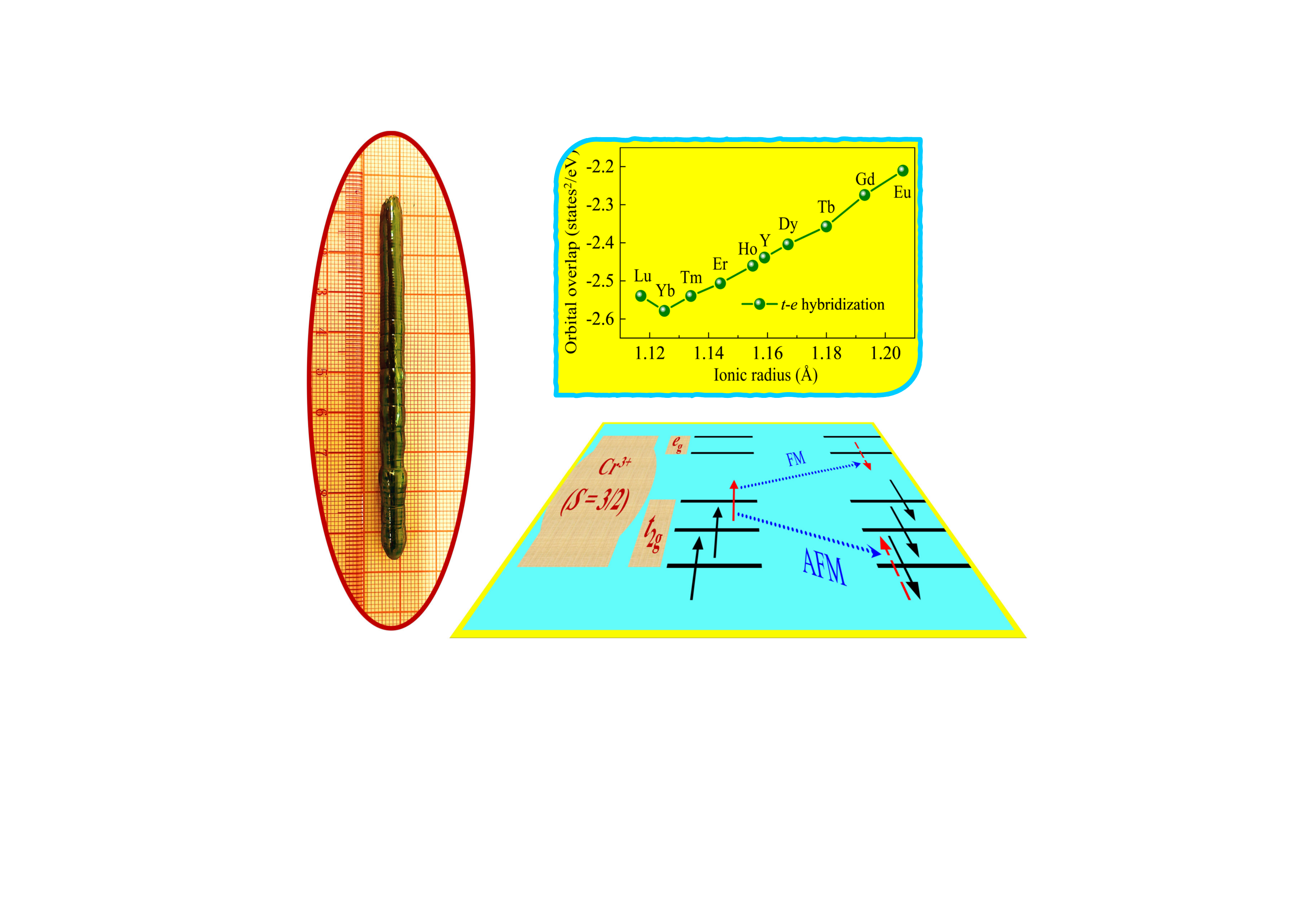}
\end{graphicalabstract}

\begin{highlights}
\item The long-standing lack of large and good-quality single crystals of orthochromates has been solved.
\item Nature of the weak ferromagnetism in a main antiferromagnetic matrix has been revealed.
\item $t$-$e$ orbital hybridization is the microscopic origin.
\item Relationship between microscopic and macroscopic properties has been correlated.
\end{highlights}


\end{frontmatter}


\section{Introduction}
\label{}

Multiferroic materials have received considerable attention due to their potential application in electric and magnetic devices \citep{Hur2004, Jones2014, Fiebig2016, Spaldin2019, HUANG2021116536}. Within the materials, long-range orders such as ferroelectric and magnetic coexist and may interact strongly on each other, providing multi-tunable parameters for tailoring the macroscopic functionalities. There are significant challenges for a complete understanding of the microscopic mechanisms underlying their intriguing macroscopic properties.

\begin{figure*} [!t]
\centering \includegraphics[width=0.82\textwidth]{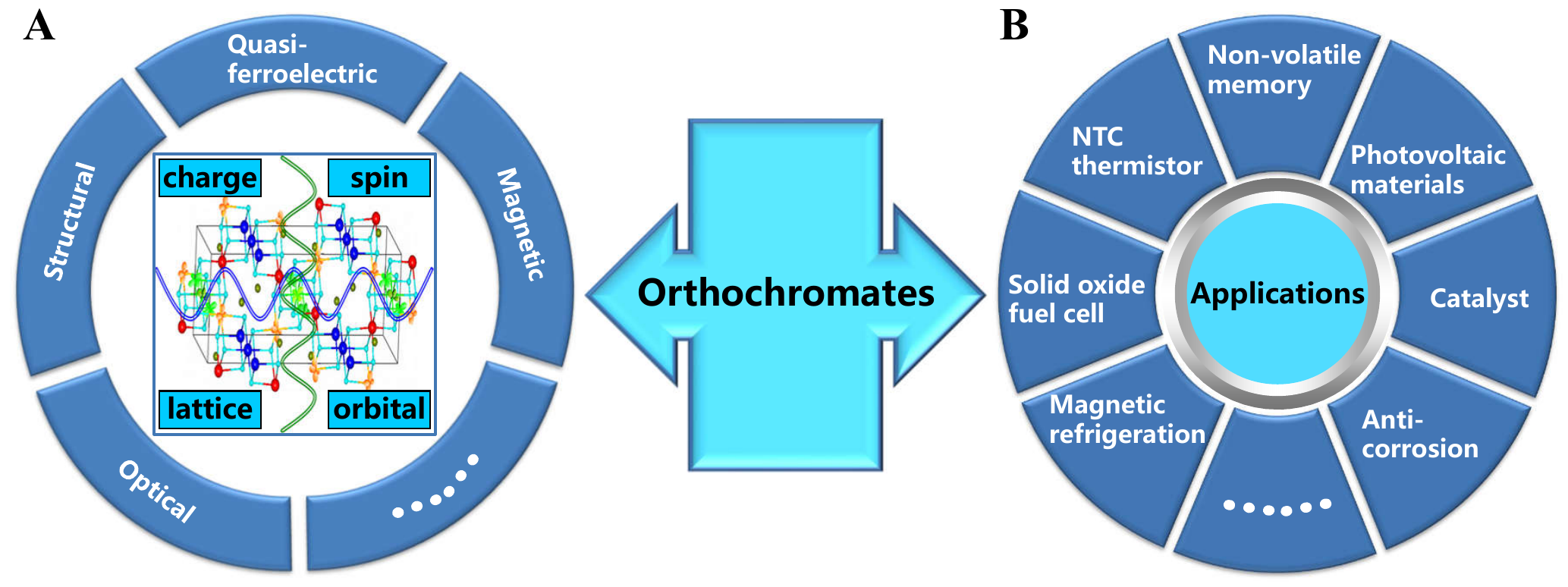}
\caption{\textbf{Properties and applications of orthochromates} \newline
(A) Orthochromates display some interesting properties such as quasiferroelectric, structural, magnetic, and optical, resulting from couplings between charge, spin, orbital, and lattice degrees of freedom. \newline
(B) These make orthochromates potential as magnetic refrigeration, solid oxide fuel cell, negative-temperature-coefficient (NTC) thermistor, non-volatile memory application, photovoltaic materials, catalyst, and anti-corrosion field.
}
\label{application}
\end{figure*}

Chromium-based perovskites have attracted considerable interests due to their intriguing magnetic and ferroelectric properties as well as potential technique applications. The set of chromium-based RECrO$_3$ (RE = rare earths) compounds was suggested to be another family of multiferroic materials, usually displaying ferroelectricity, weak ferromagnetism, and a wide application in fields of catalyst, thermistor, solid-oxide fuel cell, and non-volatile memory device (Figure~\ref{application}) \citep{Oliveira2017}. The formation of Cr$^{3+}$ magnetic orders has a strong effect on the ferroelectric property, that is, there may exist a magnetoelectric coupling effect \citep{Rajeswaran2012, Oliveira2020}. Chromium ions in RECrO$_3$ compounds hold a single-valence state, that is, Cr$^{3+}$ ($t^3_{\texttt{2g}}$$e^0_{\texttt{g}}$), naturally discarding the $e_\texttt{g}$ orbital ordering as well as its perturbation on the $t_\texttt{2g}$ electrons. This results in a tightly localized electric environment, an ideal platform for the potential coexistence of ferroelectricity and magnetism. The ferroelectric transition temperature ($T_\texttt{C} \sim$ 473 K) of YCrO$_3$ compound is much higher than that of the antiferromagnetic (AFM) phase transition of Cr$^{3+}$ ions ($T^{\texttt{Cr}}_\texttt{N} \sim$ 141.5 K) \citep{Serrao2005, Zhu2020-2}. An isosymmetric structural phase transition was observed at $\sim$ 900 K in a neutron powder diffraction study on the pulverized YCrO$_3$ single crystal, where the incompressibility of lattice constants \emph{a}, \emph{b}, and \emph{c} is anisotropic, and there exist obvious atomic displacement and charge subduction on the Y and O2 sites \citep{Zhu2020-1}. Net electric polarization was observed for polycrystalline LuCrO$_3$ and ErCrO$_3$ samples below $T^\texttt{Cr}_\texttt{N}$, indicating the presence of a possible ferroelectric state, whereas this is clearly absent above $T^\texttt{Cr}_\texttt{N}$. Most importantly, the study demonstrates that the paramagnetic (PM) nature of RE sites is not necessary to accommodate the ferroelectricity in orthochromates \citep{Meher2014}. Furthermore, the polarizations attain maximum values of $\sim 90$ ${\mu}$C/m$^2$ (at \emph{E} = 165 kV/cm for LuCrO$_3$ compound) and $\sim$ 70 $\mu$C/m$^2$ (at \emph{E} = 174 kV/cm for ErCrO$_3$ compound). Both polarizations are reversal and can be explained by either the Cr$^{3+}$ off-centring or the ferroelasticity, or their couplings, and even by Cr$^{3+}$ vacancies \citep{Meher2014}. The SmCrO$_3$ compound demonstrates an electric polarization with the maximum value of ${\sim}$ 8 ${\mu}$C/m$^2$ at \emph{E} = 1.43 kV/cm and $\sim$ 15 K \citep{Amrani2014}, which was ascribed to a breaking of the local symmetry via Cr$^{3+}$ off-centering \citep{Amrani2014}. It was observed that electric dipoles exist in DyCrO$_3$ compound, which was attributed to the displacement of Cr$^{3+}$ cations \citep{Yin2018}. Electric polarizations were observed in TbCrO$_3$ and TmCrO$_3$ compounds at \emph{E} = 1.43 kV/cm below $T^\texttt{Cr}_\texttt{N}$ \citep{Rajeswaran2012}, while the existence of electric dipoles in TmCrO$_3$ compound remains a debate \citep{Yoshii2019}.

Most previous studies focused on polycrystals, nanocrystals, and thin films. Single crystal growths of rare-earth orthochromates utilized mainly two methods: (i) one is the flux method \citep{Yin2015}. Unfortunately, impurities from the flux exist in the grown single crystals and have strong effects on macroscopic properties of the host \citep{Cooke1974,Zhu2019-1,Zhu2020-3}. In addition, single crystals grown by the flux method are small, normally millimeter in size, and not suitable for some studies that make excessive demands in sample's quality and mass. (ii) The other is the floating-zone (FZ) method with a mirror furnace. Unfortunately, the intense volatility of Cr element due to its high vapor pressure at melting points of orthochromates \citep{Philipp2003} can practically reduce the heating power of the mirror furnace \citep{Li2008-1}. This makes it not easy to stably grow the single crystals of orthochromates. As a consequence, the lack of large single crystals has been a long-standing obstacle for studying their intrinsic properties and realizing some potential applications of orthochromates.

One of the long-standing unsolved issues existing in the family of RECrO$_3$ orthochromates is the microscopic origin of the weak ferromagnetism. This is introduced by the canted AFM structure. The Cr$^{3+}$ state ($t^3_\texttt{2g}$$e^0_\texttt{g}$) in orthochromates enables a virtual charge transfer (VCT) of Cr$^{3+}$($t_\texttt{2g$\uparrow$}^3$)-O$^{2-}$-Cr$^{3+}$($e_\texttt{g}^0$) and a possible ferromagnetic (FM) competition with superexchange interactions $J \sim b^2/U$ \citep{Landron2008}, where $b$ is the overlapping integral, and $U$ denotes the on-site Coulomb interactions \citep{Slater1954, Zhou2010}. The overall Cr-O-Cr superexchange comprises two major contributions: One is from the $t_\textrm{2g}$-$t_\textrm{2g}$ hopping, producing an AFM component; the other is from the $t_\textrm{2g}$-$e_\textrm{g}$ orbital hybridization, generating FM couplings. The hybridization strength depends on the lattice distortion \citep{Siddiqu2021}, on-site Coulomb interaction \citep{Besbes2019, Yekta2021}, and $t_\textrm{2g}$-$e_\textrm{g}$ crystal field splitting \citep{Ko2007}. The spin-orbit coupling induced Dzyaloshinskii-Moriya (DM) exchange would favor a magnetic structure with spins perpendicular to each other to minimize the total energy of the system \citep{Coffey1992, Dmitrienko2014}. To describe the effect of RE sites on the Cr$^{3+}$-O$^{2-}$-Cr$^{3+}$ superexchange interactions, one can utilize density functional theory (DFT) based first-principles calculations to optimize the atomic information and correlate it with the superexchange interactions in RECrO$_3$ compounds. This provides a quantitative description of the $t$-$e$ hybridization process. Additionally, the AFM transition temperatures can be further calculated for a qualitative comparison to those measured from the grown single crystals.

In this paper, we report on a successful single crystal growth of the family of RECrO$_3$ compounds by a laser-diode FZ technique. Large single crystals with centimeter in size were obtained with the largest mass $>$ 10 g. The results from in-house characterizations on grown single crystals are in agreement with those from our first-principles calculations. Our research reveals that the $t$-$e$ hybridization process can be tuned by RE-ions and induce FM interactions in the main AFM matrix and sheds light on the coexistence of weak ferromagnetism with antiferromagnetism and ferroelectricity in orthochromates.

\begin{figure*} [!t]
\centering \includegraphics[width=0.82\textwidth]{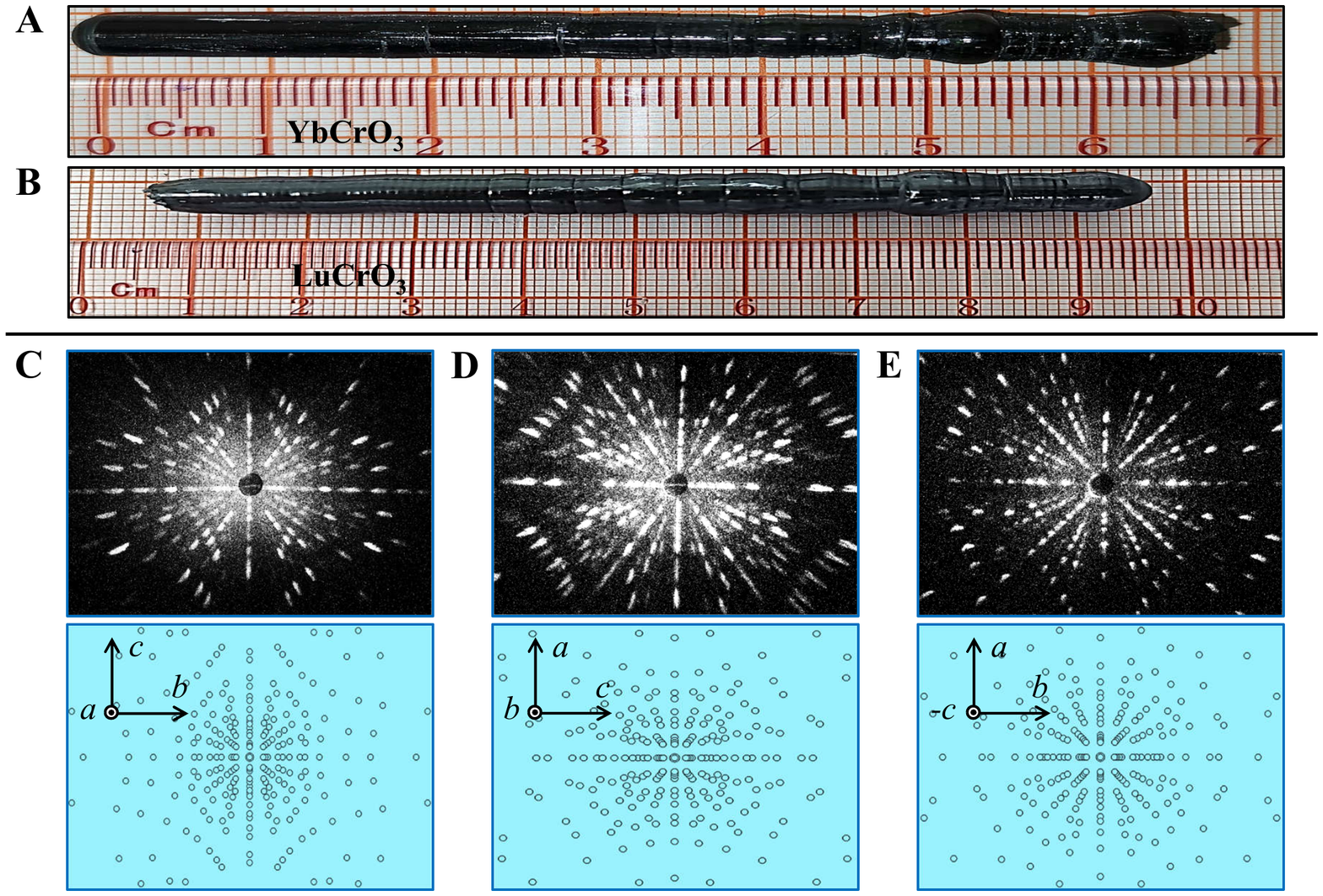}
\caption{\textbf{As-grown single crystals and neutron Laue diffraction patterns} \newline
(A) Photograph of a YbCrO$_3$ single crystal. \newline
(B) Photograph of a LuCrO$_3$ single crystal as grown by a laser-diode FZ furnace. \newline
(C-E) Neutron Laue patterns of single-crystal YCrO$_3$ (top panels) and the corresponding theoretical simulations (bottom panels). The real-space lattice vectors are marked in the bottom panels, and the crystallographic \emph{a} axis (C), \emph{b} axis (D), and \emph{c} axis (E) are perpendicular to the paper.
}
\label{SCphotos-Laue}
\end{figure*}

\begin{figure*}[!t]
\centering \includegraphics[width=0.76\textwidth]{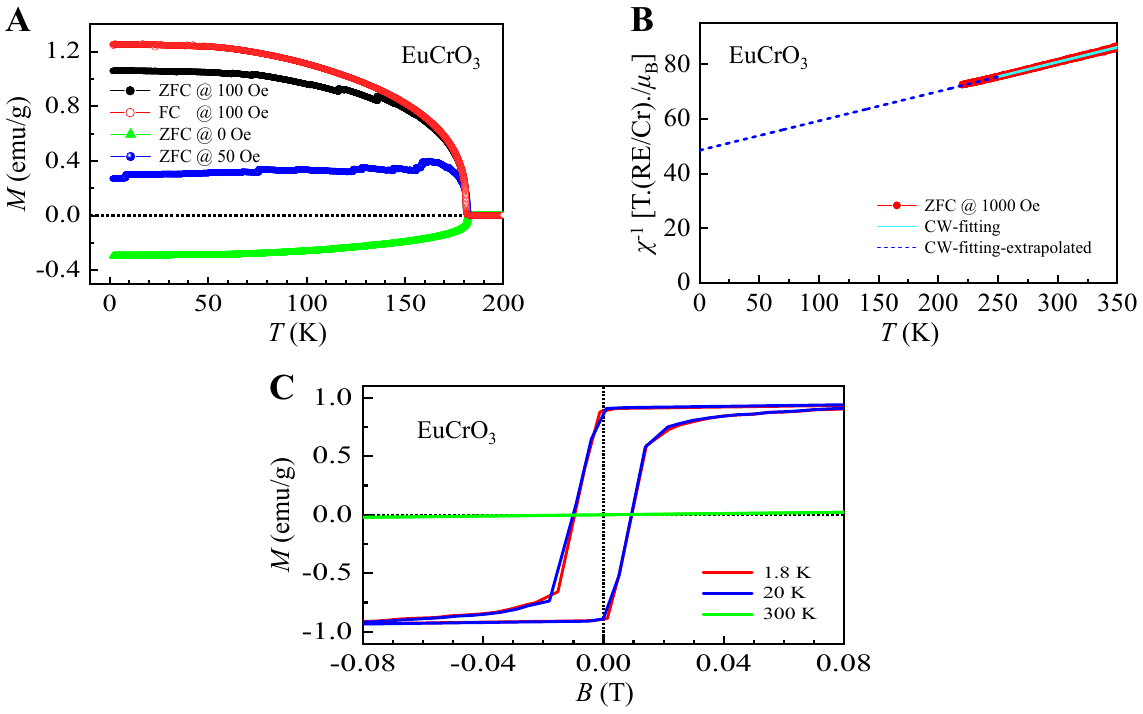}
\caption{\textbf{Magnetic property of EuCrO$_3$} \newline
(A) Magnetization as a function of temperature from 1.8--200 K measured at 0, 50, and 100 Oe. We performed both zero-field and field-cooling measurements at 100 Oe. \newline
(B) Zero-field cooling inverse magnetic susceptibility ${\chi}^{-1}$ (solid circles) at an applied magnetic field of 1000 Oe as a function of temperature in the range of 220--350 K. The solid lines represent the fits with a Curie-Weiss law from 250--350 K. The Curie-Weiss law fits were extrapolated to low temperatures shown as short-dashed lines. \newline
(C) Magnetic hysteresis loops measured at low applied magnetic fields with selected temperatures.
}
\label{MEuCO}
\end{figure*}

\begin{figure*}[!t]
\centering \includegraphics[width=0.76\textwidth]{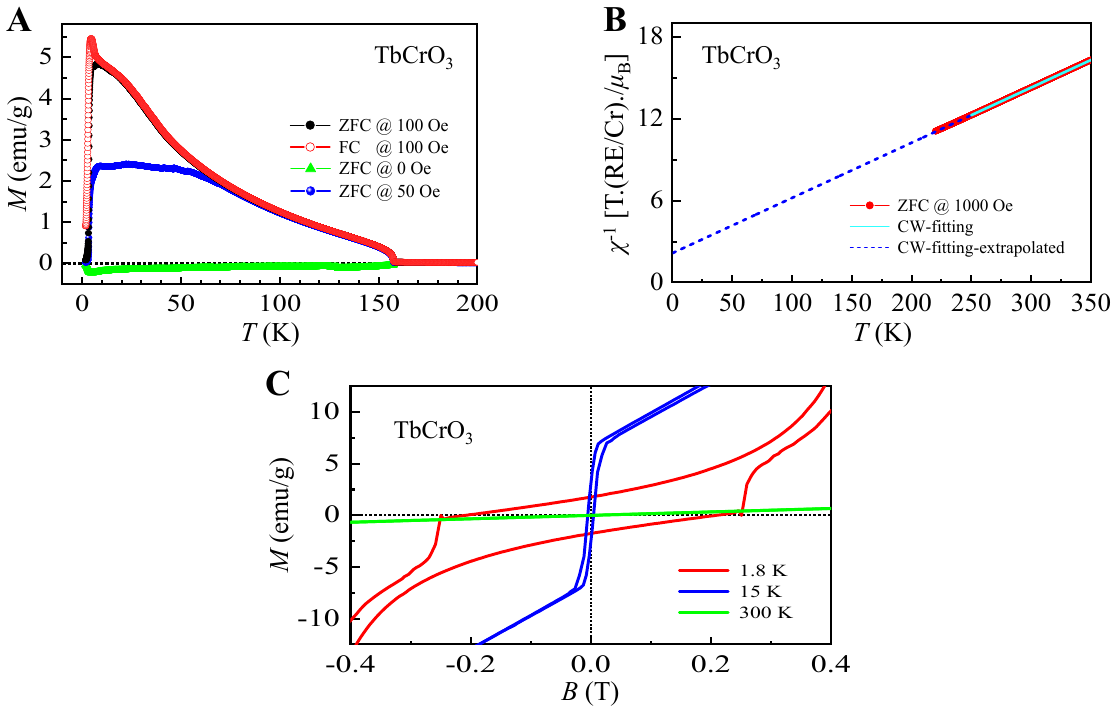}
\caption{\textbf{Magnetic property of TbCrO$_3$} \newline
(A) Magnetization as a function of temperature from 1.8--200 K measured at 0, 50, and 100 Oe. We performed both zero-field and field-cooling measurements at 100 Oe. \newline
(B) Zero-field cooling inverse magnetic susceptibility ${\chi}^{-1}$ (solid circles) at an applied magnetic field of 1000 Oe as a function of temperature in the range of 220--350 K. The solid lines represent the fits with a Curie-Weiss law from 250--350 K. The Curie-Weiss law fits were extrapolated to low temperatures shown as short-dashed lines. \newline
(C) Magnetic hysteresis loops measured at low applied magnetic fields with selected temperatures.
}
\label{MTbCO}
\end{figure*}

\begin{figure*}[!t]
\centering \includegraphics[width=0.76\textwidth]{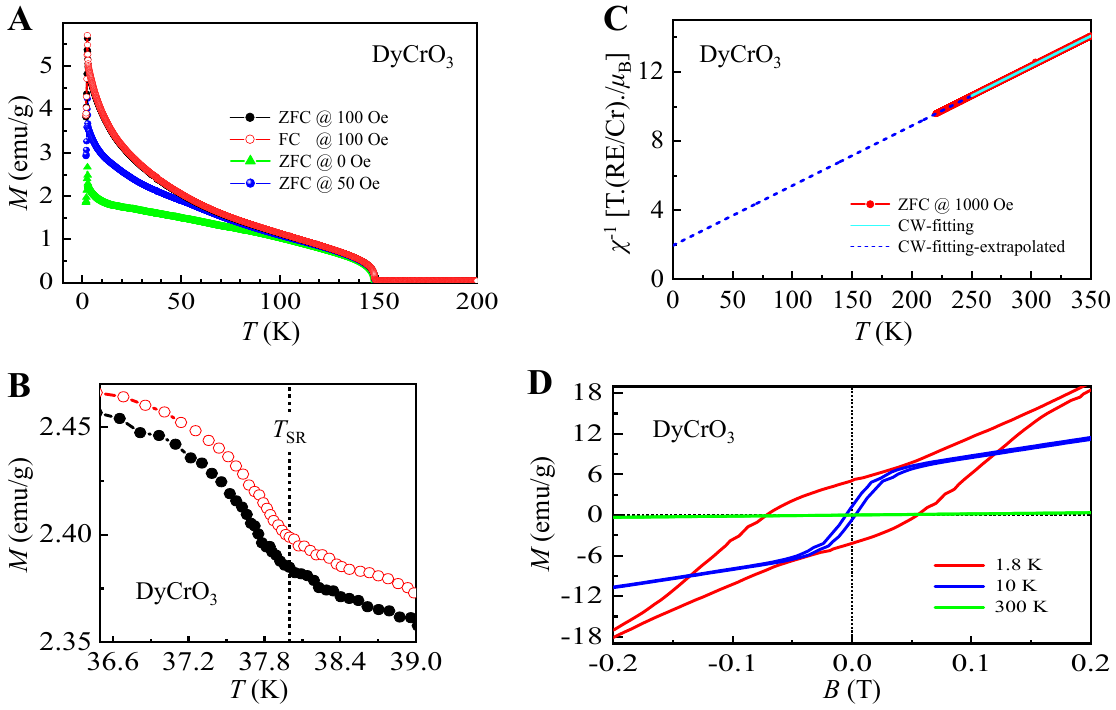}
\caption{\textbf{Magnetic property of DyCrO$_3$} \newline
(A) Magnetization as a function of temperature from 1.8--200 K measured at 0, 50, and 100 Oe. We performed both zero-field and field-cooling measurements at 100 Oe. \newline
(B) Schematically showing the determination of the spin-reorientation temperature $T_{\textrm{SR}}$. \newline
(C) Zero-field cooling inverse magnetic susceptibility ${\chi}^{-1}$ (solid circles) at an applied magnetic field of 1000 Oe as a function of temperature in the range of 220--350 K. The solid lines represent the fits with a Curie-Weiss law from 250--350 K. The Curie-Weiss law fits were extrapolated to low temperatures shown as short-dashed lines. \newline
(D) Magnetic hysteresis loops measured at low applied magnetic fields with selected temperatures.
}
\label{MDyCO}
\end{figure*}

\begin{figure*}[!t]
\centering \includegraphics[width=0.76\textwidth]{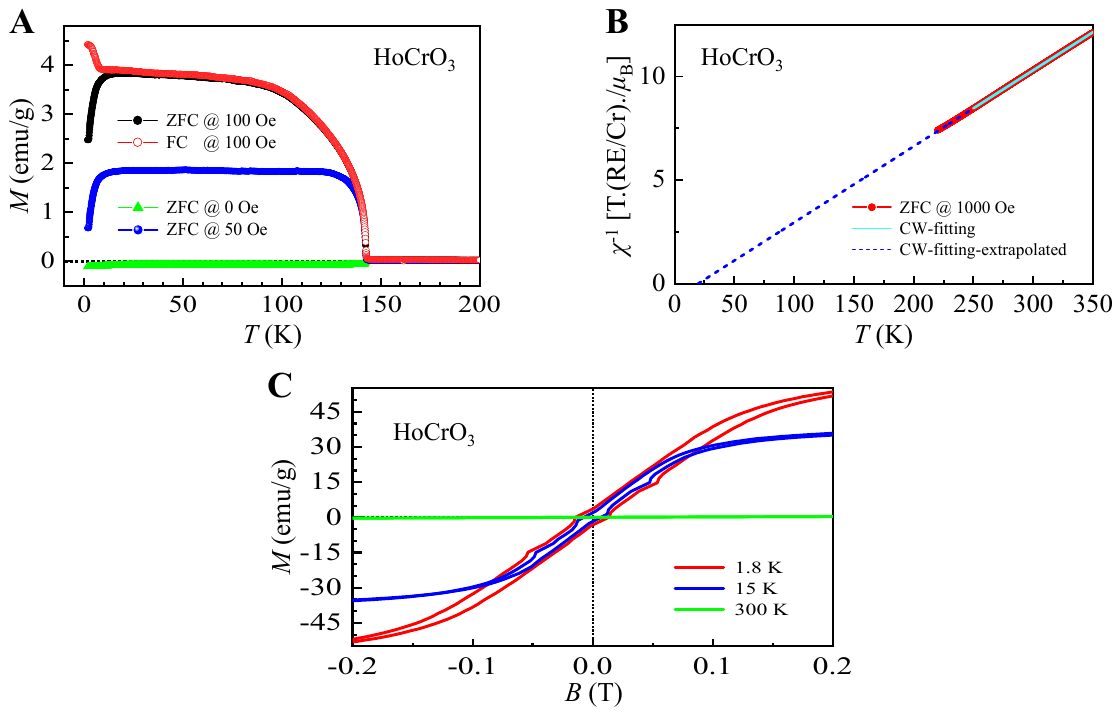}
\caption{\textbf{Magnetic property of HoCrO$_3$} \newline
(A) Magnetization as a function of temperature from 1.8--200 K measured at 0, 50, and 100 Oe. We performed both zero-field and field-cooling measurements at 100 Oe. \newline
(B) Zero-field cooling inverse magnetic susceptibility ${\chi}^{-1}$ (solid circles) at an applied magnetic field of 1000 Oe as a function of temperature in the range of 220--350 K. The solid lines represent the fits with a Curie-Weiss law from 250--350 K. The Curie-Weiss law fits were extrapolated to low temperatures shown as short-dashed lines. \newline
(C) Magnetic hysteresis loops measured at low applied magnetic fields with selected temperatures.
}
\label{MHoCO}
\end{figure*}

\begin{figure*}[!t]
\centering \includegraphics[width=0.76\textwidth]{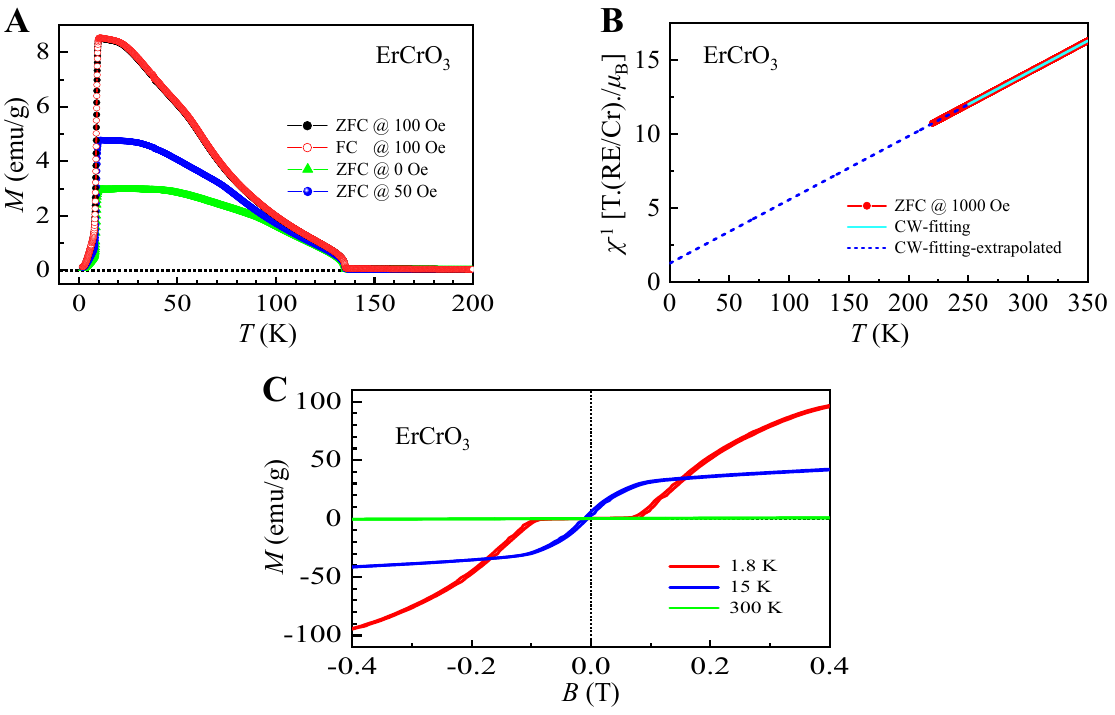}
\caption{\textbf{Magnetic property of ErCrO$_3$} \newline
(A) Magnetization as a function of temperature from 1.8--200 K measured at 0, 50, and 100 Oe. We performed both zero-field and field-cooling measurements at 100 Oe. \newline
(B) Zero-field cooling inverse magnetic susceptibility ${\chi}^{-1}$ (solid circles) at an applied magnetic field of 1000 Oe as a function of temperature in the range of 220--350 K. The solid lines represent the fits with a Curie-Weiss law from 250--350 K. The Curie-Weiss law fits were extrapolated to low temperatures shown as short-dashed lines. \newline
(C) Magnetic hysteresis loops measured at low applied magnetic fields with selected temperatures.
}
\label{MErCO}
\end{figure*}

\begin{figure*}[!t]
\centering \includegraphics[width=0.76\textwidth]{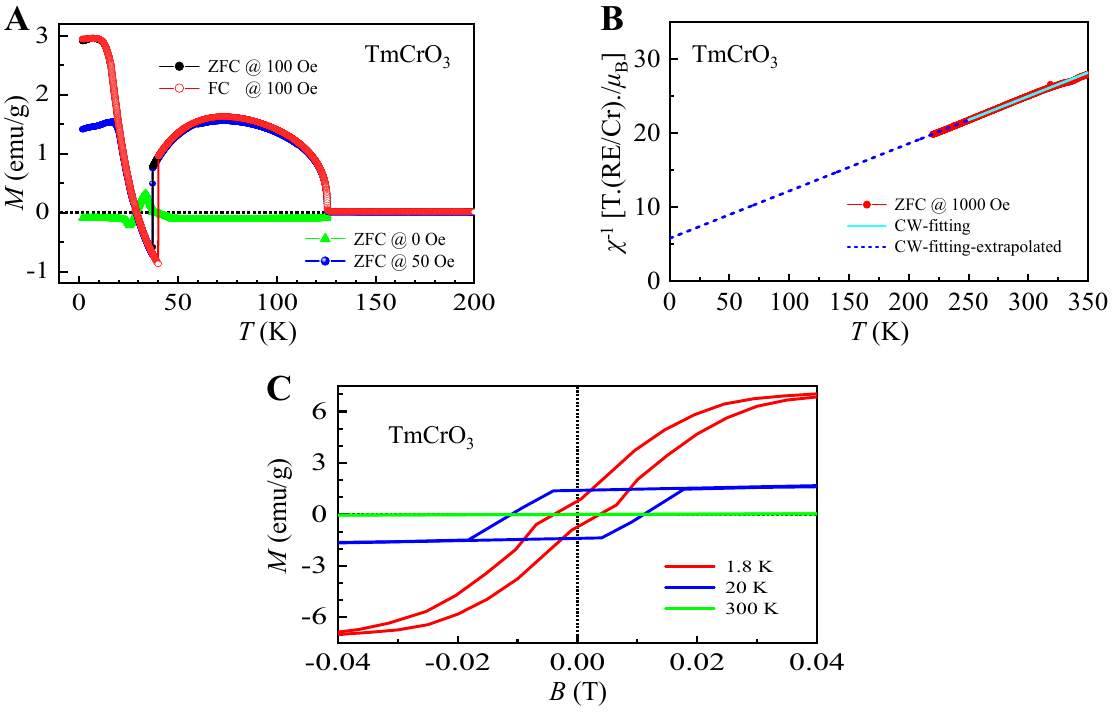}
\caption{\textbf{Magnetic property of TmCrO$_3$} \newline
(A) Magnetization as a function of temperature from 1.8--200 K measured at 0, 50, and 100 Oe. We performed both zero-field and field-cooling measurements at 100 Oe. \newline
(B) Zero-field cooling inverse magnetic susceptibility ${\chi}^{-1}$ (solid circles) at an applied magnetic field of 1000 Oe as a function of temperature in the range of 220--350 K. The solid lines represent the fits with a Curie-Weiss law from 250--350 K. The Curie-Weiss law fits were extrapolated to low temperatures shown as short-dashed lines. \newline
(C) Magnetic hysteresis loops measured at low applied magnetic fields with selected temperatures.
}
\label{MTmCO}
\end{figure*}

\begin{figure*}[!t]
\centering \includegraphics[width=0.76\textwidth]{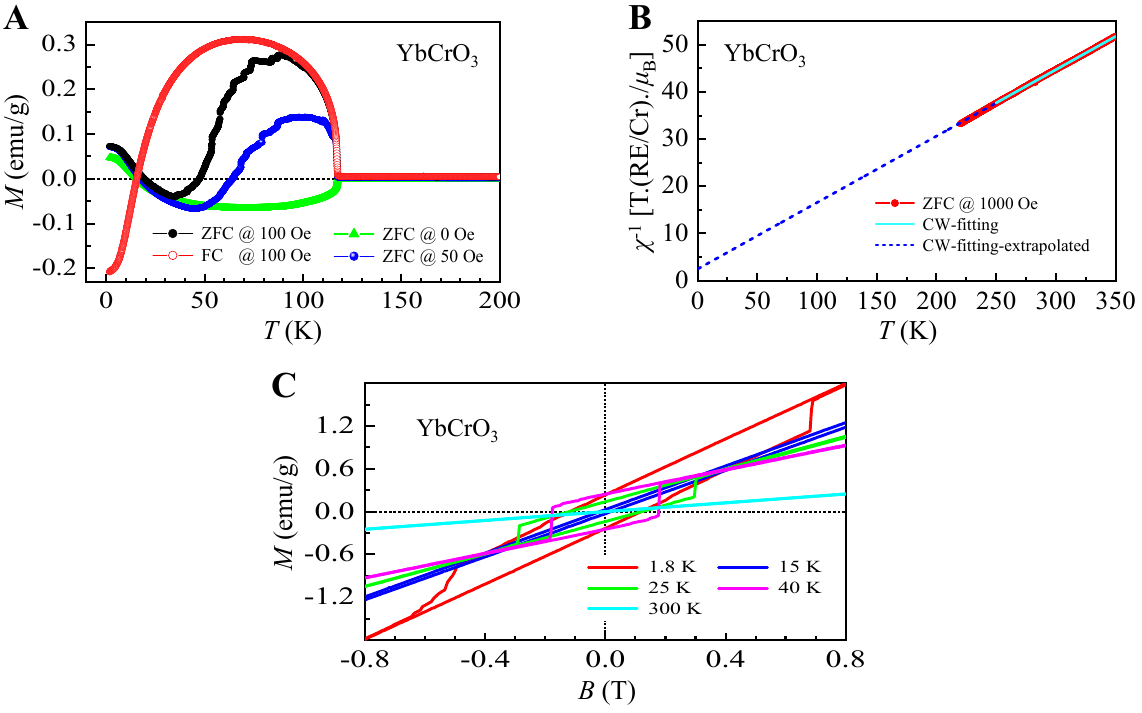}
\caption{\textbf{Magnetic property of YbCrO$_3$} \newline
(A) Magnetization as a function of temperature from 1.8--200 K measured at 0, 50, and 100 Oe. We performed both zero-field and field-cooling measurements at 100 Oe. \newline
(B) Zero-field cooling inverse magnetic susceptibility ${\chi}^{-1}$ (solid circles) at an applied magnetic field of 1000 Oe as a function of temperature in the range of 220--350 K. The solid lines represent the fits with a Curie-Weiss law from 250--350 K. The Curie-Weiss law fits were extrapolated to low temperatures shown as short-dashed lines. \newline
(C) Magnetic hysteresis loops measured at low applied magnetic fields with selected temperatures.
}
\label{MYbCO}
\end{figure*}

\begin{figure*}[!t]
\centering \includegraphics[width=0.76\textwidth]{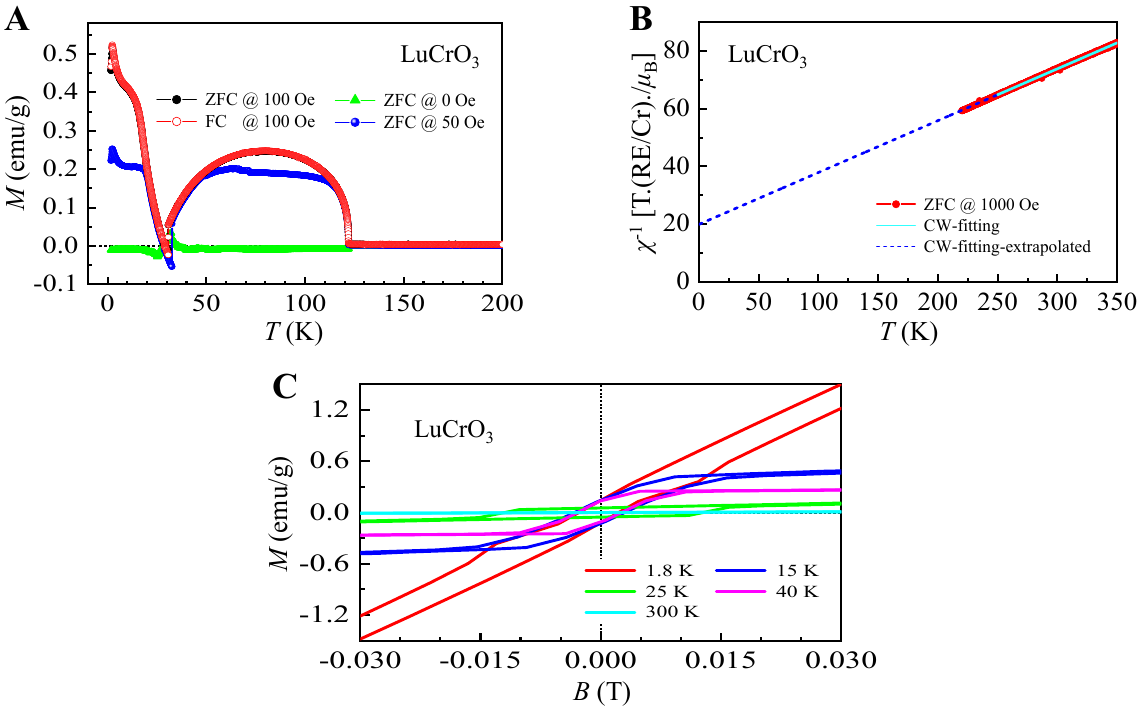}
\caption{\textbf{Magnetic property of LuCrO$_3$} \newline
(A) Magnetization as a function of temperature from 1.8--200 K measured at 0, 50, and 100 Oe. We performed both zero-field and field-cooling measurements at 100 Oe. \newline
(B) Zero-field cooling inverse magnetic susceptibility ${\chi}^{-1}$ (solid circles) at an applied magnetic field of 1000 Oe as a function of temperature in the range of 220--350 K. The solid lines represent the fits with a Curie-Weiss law from 250--350 K. The Curie-Weiss law fits were extrapolated to low temperatures shown as short-dashed lines. \newline
(C) Magnetic hysteresis loops measured at low applied magnetic fields with selected temperatures.
}
\label{MLuCO}
\end{figure*}

\begin{table*}[!t]
\small
\caption{Calculated theoretical quantum numbers of RE$^{3+}$ ions in RECrO$_3$ single crystals based on the Hund's rule: number of 4\emph{f} (4\emph{d} for Y$^{3+}$ ion) electrons, spin $S$, orbital $L$, total angular momentum $J$, Land\'{e} factors $g_J$, and the ground-state term $^{2S+1}L_J$. We listed values of the measured (meas) ($\mu_\texttt{eff-meas}$) and theoretical (theo) effective PM moments ($\mu^\texttt{Cr}_\texttt{eff-theo}$ = $g_J\mu_\texttt{B}\sqrt{S(S+1)}$ = 3.873 $\mu_\texttt{B}$ (Cr$^{3+}$: $3d^3$, $J = S = \frac{3}{2}$, $g_J =$ 2), $\mu^\texttt{RE}_\texttt{eff-theo}$ = $g_J\mu_\texttt{B}\sqrt{J(J+1)}$, and $\mu^\texttt{total}_\texttt{eff-theo}$ = $\sqrt{(\mu^\texttt{RE}_\texttt{eff-theo})^2+(\mu^\texttt{Cr}_\texttt{eff-theo})^2}$), the measured moment ($M_\texttt{meas}$ per formula at 1.8 K and 14 T), and the theoretical saturation moments ($M^\texttt{Cr}_\texttt{sat-theo}$ = $g_J\mu_\texttt{B}S$ = 3 $\mu_\texttt{B}$, $M^\texttt{RE}_\texttt{sat-theo}$ = $g_J\mu_\texttt{B}J$, and $M^\texttt{total}_\texttt{sat-theo}$ = $\sqrt{(M^\texttt{RE}_\texttt{sat-theo})^2+(M^\texttt{Cr}_\texttt{sat-theo})^2}$). The values of $T_\texttt{N}$, $\mu_\texttt{eff-meas}$, and CW temperature ($\theta_\texttt{CW}$) were extracted from the \emph{M}-\emph{T} measurements by the CW-law fitting. The values of $M_\texttt{meas}$ were extracted from the \emph{M}-\emph{B} data. The numbers in parenthesis are the estimated standard deviations of the (next) last significant digit. Some parameters of YCrO$_3$ \citep{Zhu2020-2} and GdCrO$_3$ \citep{Zhu2020-3} were referred to our previous studies.}
\label{CW-parameter}
\setlength{\tabcolsep}{6.8mm}{}
\renewcommand{\arraystretch}{1.05}
\begin{tabular} {rlllll}
\hline
\hline
\multicolumn{6}{c}{RECrO$_3$ single crystals grown by the FZ method}                                                                                                                          \\
\hline
RE$^{3+}$ =                                                             & Y$^{3+}$                       & Eu$^{3+}$   & Gd$^{3+}$                   & Tb$^{3+}$    & Dy$^{3+}$               \\
\hline
4$d^\texttt{n}$ ion                                                     & 0                              &             &                             &              &                         \\
4$f^\texttt{n}$ ions                                                    &                                & 6           & 7                           & 8            &9                        \\
$S$                                                                     & 0                              & 3           & 7/2                         & 3            &5/2                      \\
$L$                                                                     & 0                              & 3           & 0                           & 3            &5                        \\
$J$                                                                     & 0                              & 0           & 7/2                         & 6            &15/2                     \\
$g_J$                                                                   & --                             & --          & 2                           & 1.5          & 1.33                    \\
$^{2S+1}L_J$                                                            & $^{1}S_0$                      & $^{7}F_0$   & $^{8}S_{7/2}$               & $^{7}F_6$    & $^{6}H_{15/2}$          \\
\hline
$\mu_\texttt{eff-meas}$ ($\mu_\texttt{B}$)                              & 3.95                           & 6.44        & 8.40                        & 10.51        & 11.35                   \\
$\mu^\texttt{RE}_\texttt{eff-theo}$ ($\mu_\texttt{B}$)                  & 0                              & 0           & 7.937                       & 9.721        & 10.646                  \\
$\mu^\texttt{total}_\texttt{eff-theo}$ ($\mu_\texttt{B}$)               & 3.873                          & 3.873       & 8.832                       & 10.464       & 11.328                  \\ [2pt]
\hline
$M_\texttt{meas}$ ($\mu_\texttt{B}$)                                    & $\sim0.147$                    & 0.245(1)    & 6.43                        & 6.248(1)     & 4.834(1)                \\
$M^\texttt{RE}_\texttt{sat-theo}$ ($\mu_\texttt{B}$)                    & 0                              & 0           & 7                           & 9            & 10                      \\
$M^\texttt{total}_\texttt{sat-theo}$ ($\mu_\texttt{B}$)                 & 3                              & 3           & 7.616                       & 9.487        & 10.440                  \\ [2pt]
\hline
$T_\texttt{N}$ (K)                                                      & 141.5(1)                       & 181.6(1)    & 169.3(1)                    & 157.9(1)     & 148.5(1)                \\
$\theta_\texttt{CW}$ (K)                                                & $-$433.2(6)                    & $-$450.4(15)& $-$20.33(4)                 & $-$53.3(1)   & $-$56.5(1)              \\
\hline
\hline
\multicolumn{6}{c}{continued}                                                                                                                                                                 \\
\hline
\hline
RE$^{3+}$ =                                                             & Ho$^{3+}$                      & Er$^{3+}$   & Tm$^{3+}$                   & Yb$^{3+}$    & Lu$^{3+}$               \\
\hline
4$f^\texttt{n}$ ion                                                     & 10                             & 11          & 12                          & 13           & 14                      \\
$S$                                                                     & 2                              & 3/2         & 1                           & 1/2          & 0                       \\
$L$                                                                     & 6                              & 6           & 5                           & 3            & 0                       \\
$J$                                                                     & 8                              & 15/2        & 6                           & 7/2          & 0                       \\
$g_J$                                                                   & 1.25                           & 1.2         & 1.167                       & 1.143        & --                      \\
$^{2S+1}L_J$                                                            & $^5I_8$                        & $^4I_{15/2}$ &$^3H_6$                     & $^2F_{7/2}$  & $^1S_0$                 \\
\hline
$\mu_\texttt{eff-meas}$ ($\mu_\texttt{B}$)                              & 11.03                          & 10.20       & 8.35                        & 5.63         & 4.98                    \\
$\mu^\texttt{RE}_\texttt{eff-theo}$ ($\mu_\texttt{B}$)                  & 10.607                         & 9.581       & 7.561                       & 4.536        & 0                       \\
$\mu^\texttt{total}_\texttt{eff-theo}$ ($\mu_\texttt{B}$)               & 11.292                         & 10.334      & 8.495                       & 5.964        & 3.873                   \\ [2pt]
\hline
$M_\texttt{meas}$ ($\mu_\texttt{B}$)                                    & 3.999(1)                       & 6.385(1)    & 2.868(1)                    & 0.883(1)     & 1.197(1)                \\
$M^\texttt{RE}_\texttt{sat-theo}$ ($\mu_\texttt{B}$)                    & 10                             & 9           & 7                           & 4            & 0                       \\
$M^\texttt{total}_\texttt{sat-theo}$ ($\mu_\texttt{B}$)                 & 10.440                         & 9.487       & 7.616                       & 5            & 3                       \\ [2pt]
\hline
$T_\texttt{N}$ (K)                                                      & 143.2(1)                       & 135.4(1)    & 125.9(1)                    & 117.9(1)     & 122.3(1)                \\
$\theta_\texttt{CW}$ (K)                                                & 19.6(1)                        & $-$30.0(1)  & $-$90.6(4)                  & $-$17.4(1)   & $-$110.4(2)             \\
\hline
\hline
\end{tabular}
\end{table*}

\begin{table}[t!]
\small
\caption{Summary of the previously reported magnetic structures of RE$^{3+}$ and Cr$^{3+}$ ions in RECrO$_3$ (RE = Gd, Tb, Dy, Ho, Er, Tm, Yb, Lu) compounds. The crystal structure is orthorhombic with $Pbnm$ space group (No. 62). Here $T_\textrm{N}^\textrm{RE}$ = AFM transition temperature of RE$^{3+}$ ions; SR = spin reorientation; $T_\textrm{SR}^\textrm{Cr}$ = SR temperature of Cr$^{3+}$ ions; $T_\textrm{N}^\textrm{Cr}$ = AFM transitions temperature of Cr$^{3+}$ ions. The check mark ($\surd$) represents that there exists a magnetic SR phase transition. These were summarized from the following references: GdCrO$_3$ \citep{Cooke1974}, TbCrO$_3$ \citep{Gordon1976}, DyCrO$_3$ \citep{Bertaut1963, Tsushima1974, Krynetskii1997}, HoCrO$_3$ \citep{Shamir1981, Su2011}, ErCrO$_3$ \citep{Hornreich1978, Shamir1981}, TmCrO$_3$ \citep{Shamir1981}, YbCrO$_3$ \citep{Shamir1981}, and LuCrO$_3$ \citep{Shamir1981}.}
\label{Magentic-structures}
\setlength{\tabcolsep}{9.85mm}{}
\renewcommand{\arraystretch}{1.05}
\begin{tabular}{lcccc}
\hline
\hline
RECrO$_3$     & RE$^{3+}$                            & Cr$^{3+}$                             & SR              & Cr$^{3+}$                                                         \\ [2pt]
              & ($T < T_\textrm{N}^\textrm{RE}$)     & ($T < T_\textrm{SR}^\textrm{Cr}$)     & (Cr$^{3+}$)     & $(T_\textrm{SR}^\textrm{Cr} < T < T_\textrm{N}^\textrm{Cr})$      \\ [2pt]
\hline
GdCrO$_3$     & $F_xC_y$                             & $F_xG_z$                              & $\surd$         & $G_xF_z$                                                          \\
TbCrO$_3$     & $F_xC_y$                             & $-$                                   & $-$             & $F_xG_z$                                                          \\
DyCrO$_3$     & $G_xA_y$                             & $-$                                   & $-$             & $F_xG_z$                                                          \\
HoCrO$_3$     & $-F_x$$-C_y$                         & $-$                                   & $-$             & $G_z$                                                             \\
ErCrO$_3$     & $-C_z$                               & $G_y$                                 & $\surd$         & $G_xF_z$                                                          \\
TmCrO$_3$     & $-$                                  & $-$                                   & $-$             & $F_xG_z$                                                          \\
YbCrO$_3$     & $-$                                  & $-$                                   & $-$             & $F_xG_z$                                                          \\
LuCrO$_3$     & $-$                                  & $-$                                   & $-$             & $F_xG_z$                                                          \\
\hline
\hline
\end{tabular}
\end{table}

\section{Results}

\subsection{Grown single crystals}

Single-crystal materials hold translational symmetry of long-range building blocks, therefore, they provide reliable information of structures and properties of matters \citep{li2018millimeter, sun2019anisotropic, xiong2021near, CHENG20171039, ZHANG20201694}. Exploring and optimizing the single crystal growth parameters are time-consuming and labor-intensive processes \citep{Li2008-1}. We have for the first time grown large single crystals of the family of RECrO$_3$ (RE = Y, Eu, Gd, Tb, Dy, Ho, Er, Tm, Yb, and Lu) compounds. Photos of some representative single crystals as-grown were exhibited in Figures~\ref{SCphotos-Laue}A and \ref{SCphotos-Laue}B, where the YbCrO$_3$ (Figure~\ref{SCphotos-Laue}A, $\sim$ 7 cm in length) and LuCrO$_3$ (Figure~\ref{SCphotos-Laue}B, $\sim$ 10 cm) crystals have a diameter of $\phi = 6$--8 mm and very shining surfaces. So far, the largest single crystal of orthochromates we have grown is $>$ 10 g for EuCrO$_3$ compound. Our crystal growth engineering of RECrO$_3$ compounds produced a China invention patent \citep{Zhu2019-1}.

\subsection{Neutron Laue diffraction patterns}

We investigated the grown single crystals with a neutron Laue diffraction study. The top panels of Figures~\ref{SCphotos-Laue}C-\ref{SCphotos-Laue}E show the monitored neutron Laue patterns of a YCrO$_3$ single crystal with the three axes perpendicular to the paper: \emph{a}-axis (Figure~\ref{SCphotos-Laue}C), \emph{b}-axis (Figure~\ref{SCphotos-Laue}D), and \emph{c}-axis (Figure~\ref{SCphotos-Laue}E). All patterns display symmetric and very strong diffraction spots, indicating a good quality of the grown single crystal. As shown in the bottom panels of Figures~\ref{SCphotos-Laue}C-\ref{SCphotos-Laue}E, we simulated theoretically the three patterns with the software of OrientExpress \citep{Ouladdiaf2006}, which further confirms the good quality of our single crystals.

\subsection{Magnetic properties}

To clearly show features of dc magnetization in the vicinity of magnetic phase transitions, we present temperature-dependent data in the temperature range of 1.8--200 K. For the whole-temperature data (1.8--400 K), one can refer to Figure~S1. We measured the zero-field cooling (ZFC) magnetization at zero applied magnetic field, trying to explore the negative magnetization behavior and address whether the observed weak ferromagnetism is spontaneous.

We measured the ZFC magnetization data at 1000 Oe in the temperature range of 250--350 K and calculated the inverse magnetic susceptibility $\chi^{-1} = B/M$, which can be well fit with the Curie-Weiss (CW) law for a pure paramagnetic (PM) state \citep{Li2006, Li2007_2, Li2009, Zhu2020-1},
\begin{eqnarray}
\chi^{-1}(T) = \frac{3k_B(T - \Theta_{\texttt{CW}})}{N_A \mu^2_{\texttt{eff}}},
\label{CWLaw}
\end{eqnarray}
where $k_B =$ 1.38062 $\times$ 10$^{-23}$ J/K is the Boltzmann constant, ${\Theta}_\texttt{CW}$ is the PM CW temperature, $N_A =$ 6.022 $\times$ 10$^{23}$ mol$^{-1}$ is Avogadro's constant, and $\mu_{\texttt{eff}}$ is the effective PM moment. We fit the data to Equation~(\ref{CWLaw}), extracting experimental values of the AFM transition temperature $T^{\texttt{Cr}}_\texttt{N}$, ${\Theta}_\texttt{CW}$, and $\mu_{\texttt{eff-meas}}$ of RECrO$_3$ single crystals. These measured values were listed in Table~\ref{CW-parameter}. For the RECrO$_3$ compounds, it is pointed out that both RE$^{3+}$ and Cr$^{3+}$ ions contribute to the effective PM moment, therefore, theoretical total effective PM moment $\mu^\texttt{total}_\texttt{eff-theo} = \sqrt{(\mu^\texttt{RE}_\texttt{eff-theo})^2 + (\mu^\texttt{Cr}_\texttt{eff-theo})^2}$, where $\mu^\texttt{RE}_\texttt{eff-theo} =$ $g\mu_\texttt{B} \sqrt{J(J + 1})$, and $\mu^\texttt{Cr}_\texttt{eff-theo} =$ $g\mu_\texttt{B} \sqrt{S(S + 1}) =$ 3.873 ${\mu}_\texttt{B}$ (where $S =$ 3/2). The quantum numbers of RE$^{3+}$ and Cr$^{3+}$ ions and the calculated values of $\mu^\texttt{RE}_\texttt{eff-theo}$ and $\mu^\texttt{total}_\texttt{eff-theo}$ were listed in Table~\ref{CW-parameter} for a comparison to the corresponding experimental values. Within 250--350 K, the fits coincide well with the measured magnetization of RECrO$_3$ (RE = Eu, Tb, Dy, Ho, Er, Tm, Yb, and Lu) single crystals. As temperature decreases from 250 to 220 K, the measured data did not show obvious deviation from the Curie-Weiss law fitting (short-dashed line), which confirms the validity of our theoretical fits and that the RECrO$_3$ (RE = Eu, Tb, Dy, Ho, Er, Tm, Yb, and Lu) single crystals remain a pure PM state in the temperature range of 220--350 K.

Magnetic hysteresis loops at the low-field regimes were measured. The isothermal field dependence of magnetization with applied magnetic field from $-$14 to 14 T was supplemented in Figure~S2.

In the following, we present analyses of the magnetic properties of RECrO$_3$ (RE = Eu, Tb, Dy, Ho, Er, Tm, Yb, and Lu) single crystals one by one.

\subsubsection{EuCrO$_3$}

EuCrO$_3$ demonstrates negative magnetization behavior at zero magnetic field below $T^{\texttt{Cr}}_\texttt{N}$ (Figure~\ref{MEuCO}A), whereas the ZFC magnetization at \emph{B} = 50 Oe increases sharply below $T^{\texttt{Cr}}_\texttt{N}$ within a thermal regime of $\sim$ 1.2 K and then flattens to 1.8 K upon cooling. The field-cooling (FC) magnetization at 100 Oe increases by $\sim$ 18\% at 1.8 K. The magnetization measured at 50 and 100 Oe resembles the features of a weak FM state. We determined that $T^{\texttt{Cr}}_\texttt{N}$ = 181.6(1) K for a EuCrO$_3$ single crystal. Below $T^{\texttt{Cr}}_\texttt{N}$, EuCrO$_3$ enters a canted AFM state, probably because of the Dzyaloshinskii-Moriya interactions of Cr$^{3+}$ ions.

To avoid the effect of non-intrinsic magnetic contributions at low applied magnetic fields, we used the ZFC magnetization measured from 250 to 350 K at 1000 Oe for CW fitting (Figure~\ref{MEuCO}B). This produces an effective PM moment $\mu_\texttt{eff-meas}$ = 6.44 $\mu_\texttt{B}$, which is considerably larger than the theoretical value $\mu^\texttt{total}_\texttt{eff-theo}$ = 3.873 $\mu_\texttt{B}$ (Table~\ref{CW-parameter}) and a CW temperature $\Theta_\texttt{CW} = -$450.4(15) K.

The field dependence of magnetization measured at 1.8 K is displayed in Figure~\ref{MEuCO}C, where clear magnetic hysteresis loops are observed at 1.8 and 20 K, and there is nearly no difference between them. Remanent magnetization $M_{\texttt{r}}$ $\sim$ 0.91 emu g$^{-1}$ and coercive field $B_{\texttt{c}}$ $\sim$ 95 Oe were determined. The magnetic hysteresis loop closes at $\sim$ 867 Oe, after which the ZFC magnetization increases linearly as a function of the magnetic field, with a slope of $\frac{dM}{dB} =$ 0.322(1) emu g$^{-1}$ T$^{-1}$ (Figure~S2A). Given that the theoretical saturation magnetic moment of Cr$^{3+}$ ions is $M^\texttt{Cr}_\texttt{sat-theo} = g_J\mu_\texttt{B}S =$ 3 $\mu_\texttt{B}$, where $g_J =$ 2 and $S = \frac{3}{2}$ for pure ionic Cr$^{3+}$ ions (Table~\ref{CW-parameter}), it is inferred that attaining a full magnetic saturation state requires $B \approx$ 203 T. At 1.8 K and 14 T, a magnetization of $M_\texttt{meas}$ = 0.245(1) $\mu_\texttt{B}$/Cr was reached (Figure~S2A), which is equal to $\sim$ 8.2\% of the theoretical value $M^\texttt{total}_\texttt{sat-theo}$ = 3 $\mu_\texttt{B}$/Cr.

Our study indicates a possible existence of competition between FM and AFM exchange interactions. We did not observe the magnetic ordering of Eu$^{3+}$ ions, which is consistent with a previous study on polycrystalline EuCrO$_3$ samples \citep{Taheri2016}.

\subsubsection{TbCrO$_3$}

We determined that $T^{\texttt{Cr}}_\texttt{N}$ = 157.9(1) K for a TbCrO$_3$ single crystal (Figure~\ref{MTbCO}A). Below $T^{\texttt{Cr}}_\texttt{N}$, the ZFC magnetization at 0 Oe was negative, whereas the magnetization curves at 50 and 100 Oe increased steady after a sharp enhancement around $T^{\texttt{Cr}}_\texttt{N}$ within $\sim$ 0.56 K. Below $\sim$ 70 K, the ZFC magnetization at 100 Oe evidently exceeded that measured at 50 Oe. At $\sim$ 7.7 K, a kink appeared in the magnetization at 50 and 100 Oe (ZFC at 50 and 100 Oe: down turn; FC at 100 Oe: up turn), indicating the formation of Tb$^{3+}$ spin ordering. Below $\sim$ 4.5 K, a sharp decrease in the magnetization at 50 and 100 Oe and a sharp increase in that at 0 Oe were observed. Below $\sim$ 3 K, the ZFC magnetization at 0, 50, and 100 Oe approached zero. Therefore, we observe elaborate magnetic phase transitions with temperature in the TbCrO$_3$ single crystal.

The CW fitting result of $\mu_\texttt{eff-meas}$ = 10.51 $\mu_\texttt{B}$ (Figure~\ref{MTbCO}B) is nearly equal to the theoretical value $\mu^\texttt{total}_\texttt{eff-theo}$ = 10.464 $\mu_\texttt{B}$. The CW temperature $\Theta_\texttt{CW} = -$53.3(1) K indicates a weak competition between the FM and AFM interactions in our TbCrO$_3$ single crystal.

A clear magnetic hysteresis loop was observed with $M_{\texttt{r}}$ $\sim$ 2.75 emu g$^{-1}$ and $B_{\texttt{c}}$ $\sim$ 48 Oe at 15 K (Figure~\ref{MTbCO}C), indicating a weak FM state of the Cr$^{3+}$ ions. When $T =$ 1.8 K, both Tb$^{3+}$ and Cr$^{3+}$ spins order, and the observed magnetization loop was stretched along the \emph{B} axis and squeezed along the magnetization axis, leading to a parallelogram-shaped hysteresis loop with $M_{\texttt{r}}$ $\sim$ 1.82 emu g$^{-1}$ and $B_{\texttt{c}}$ $\sim$ 2500 Oe (Figure~\ref{MTbCO}C). Such a twisted loop has not been previously observed in either polycrystalline or single-crystalline (grown by the flux method) samples \citep{Vagadia2018, Yin2016}. This indicates a stronger coupling between the Cr$^{3+}$ and Tb$^{3+}$ spins in our TbCrO$_3$ single crystal. As \emph{B} increased, the loop at 1.8 K gets a quick saturation at $B_{\texttt{s}}$ $\sim$ 1.67 T, while that at 15 K increased smoothly and attained a plateau at $\sim$ 6.8 T. At 1.8 K and 14 T, the magnetization reached $M_\texttt{meas}$ = 6.248(1) $\mu_\texttt{B}$, which is $\sim$ 34\% smaller than the theoretical value $M^\texttt{total}_\texttt{sat-theo}$ = 9.487 $\mu_\texttt{B}$ (Table~\ref{CW-parameter}).

TbCrO$_3$ entered a long-range canted AFM state of Cr$^{3+}$ ions below $T^{\texttt{Cr}}_\texttt{N}$ = 157.9(1). The possible long-range AFM order of Tb$^{3+}$ ions was formed below $T^{\texttt{Tb}}_\texttt{N} \sim$ 7.7 K. Strong coupling was observed between the spin orders of the Cr$^{3+}$ and Tb$^{3+}$ ions. Furthermore, a weak competition between the FM and AFM interactions of the Cr$^{3+}$ ions was observed.

\subsubsection{DyCrO$_3$}

The DyCrO$_3$ single crystal underwent a magnetic phase transition from the PM state to a canted AFM phase at $T^{\texttt{Cr}}_\texttt{N}$ = 148.5(1) K (Figure~\ref{MDyCO}A), corresponding to the formation of Cr$^{3+}$ spin ordering. A kink appeared at $T^{\texttt{Cr}}_\texttt{SR}$ $\sim$ 38 K in the magnetization at 0, 50, and 100 Oe (Figure~\ref{MDyCO}B), which is attributed to the spin reorientation of the Cr ions. A similar observation was previously reported for DyCrO$_3$ single crystals grown by the flux method, where the appearance of the kink was believed to be caused by the spin reorientation of Dy$^{3+}$ ions \citep{Yin2016}. Below $T^{\texttt{Cr}}_\texttt{N}$, the magnetization increased smoothly until the onset of a sharp enhancement at $\sim$ 17 K, reaching a maximum at $\sim$ 2.54 K and then followed by a quick reduction. Therefore, we observed an AFM phase transition of the Dy$^{3+}$ ions at $T^{\texttt{Dy}}_\texttt{N}$ $\sim$ 2.8 K.

The CW fitting produces $\mu_\texttt{eff-meas}$ = 11.35 $\mu_\texttt{B}$ (Figure~\ref{MDyCO}C), which is nearly equal to the theoretical value $\mu^{\texttt{total}}_\texttt{eff-theo}$ = 11.328 $\mu_\texttt{B}$. CW temperature, $\Theta_\texttt{CW} = -$56.5(1) K (Table~\ref{CW-parameter}).

Isothermal magnetic hysteresis loops were observed at 1.8 ($M_{\texttt{r}}$ $\sim$ 4.7 emu g$^{-1}$; $B_{\texttt{c}}$ $\sim$ 639 Oe) and 10 K ($M_{\texttt{r}}$ $\sim$ 1.1 emu g$^{-1}$; $B_{\texttt{c}}$ $\sim$ 41 Oe) (Figure~\ref{MDyCO}D), confirming that Cr$^{3+}$ spins hold a canted AFM state. For the Dy$^{3+}$ sublattices, we could only conclude that the spins form an AFM state. At both 1.8 and 10 K, the magnetization curves finally flattened with $M_\texttt{meas}$ = 4.834(1) $\mu_\texttt{B}$ at 14 T (Figure~S2C). This value was $\sim$ 53.7\% less than the theoretical value of $M^\texttt{total}_\texttt{sat-theo}$ = 10.440 $\mu_\texttt{B}$ (Table~\ref{CW-parameter}).

\subsubsection{HoCrO$_3$}

We determined that $T^\texttt{Cr}_\texttt{N}$ = 143.2(1) K for a HoCrO$_3$ single crystal (Figure~\ref{MHoCO}A). Below $T^\texttt{Cr}_\texttt{N}$, the ZFC magnetization at 0 Oe was negative, whereas those at 50 and 100 Oe were positive and increased smoothly in the temperature regimes of $\sim$ 19 K (at 50 Oe) and $\sim$ 47 K (at 100 Oe) and then flatten until $T^\texttt{Ho}_\texttt{N} \sim$ 7.82 K at which Ho$^{3+}$ ions order antiferromagnetically. Below $\sim$ 2.4 K, all curves were flattened. When 2.4 K $< T < T^\texttt{Ho}_\texttt{N}$, the magnetization curves at ZFC (down turn) and FC (up turn) 100 Oe demonstrated an inverse trend; above $T^\texttt{Ho}_\texttt{N}$, they coincided. The degree of canting of the Cr$^{3+}$ AFM structure determines the strength of the resulting ferromagnetism along the \emph{c} axis. This prevents the formation of an AFM structure of the Ho$^{3+}$ ions. The difference between the ZFC and FC magnetization at 100 Oe is controlled by the competition between the Zeeman energy generated by the applied magnetic field, crystal field, AFM interaction strength of Ho$^{3+}$ ions, and magnetic anisotropy \citep{HFLi2016}.

We obtained $\mu_\texttt{eff-meas}$ = 11.03 $\mu_\texttt{B}$, which is comparable to $\mu^\texttt{total}_\texttt{eff-theo}$ = 11.292 $\mu_\texttt{B}$, and $\Theta_\texttt{CW}$ = 19.6(1) K (Figure~\ref{MHoCO}B, Table~\ref{CW-parameter}). The previous study on polycrystalline HoCrO$_3$ sample shows $\mu_\texttt{eff-meas}$ = 11.55 $\mu_\texttt{B}$ and $\Theta_\texttt{CW} = -$24.0 K \citep{Su2011}. Obtaining evidence of short-range exchange interactions and magnetic fluctuations of Ho$^{3+}$ spins reported previously by quasielastic and inelastic neutron scattering studies on HoCrO$_3$ powder samples \citep{Kumar2016, Chatterji2017} necessitates more in-house characterizations with the HoCrO$_3$ single crystal.

At 1.8 and 15 K, we observed magnetic hysteresis loops in step-increasing mode (Figure~\ref{MHoCO}C). We extracted $M_{\texttt{r}}$ $\sim$ 3.5 (1.8 K) and 1.9 (15 K) emu g$^{-1}$ and the corresponding $B_{\texttt{c}}$ $\sim$ 140 and 72 Oe, respectively. At 15 K, the magnetization increased linearly at $B \leq $ 618 Oe and then proceeded smoothly into a plateau at $\sim$ 3.2 T; in contrast, at 1.8 K and $B \leq $ 1092 Oe, the magnetization almost increased linearly with increasing magnetic field and then attained $M_\texttt{meas}$ = 3.999(1) $\mu_\texttt{B}$ at 14 T (Figure~S2D). This value was only $\sim$ 38.3\% of the theoretical value $M^\texttt{total}_\texttt{sat-theo}$ = 10.440 $\mu_\texttt{B}$ (Table~\ref{CW-parameter}).

\subsubsection{ErCrO$_3$}

Below $T^\texttt{Cr}_\texttt{N}$ = 135.4(1) K \citep{Eibschutz1970}, there was a small sharp enhancement in the magnetization within $\sim$ 0.72 K, which then increased smoothly until an onset of a sudden decrease at $T^\texttt{Cr}_\texttt{SR}$ $\sim$ 9.7 K (Figure~\ref{MErCO}A). The decrease in magnetization is attributed to the spin reorientation of Cr$^{3+}$ ions from $\Gamma_4$ ($G_x$, $A_y$, $F_z$; $F^R_z$) to $\Gamma_1$ ($A_x$, $G_y$, $C_z$; $C^R_z$) or the $\Gamma_1$ (0) spin configuration \citep{Su2010, Su2012}. Below $T^\texttt{Er}_\texttt{N}$ $\sim$ 8 K, the magnetization at 0, 50, and 100 Oe decreased linearly. No difference was found in the ZFC and FC magnetization at 100 Oe.

The CW fitting resulted in an effective PM magnetic moment of 10.20 $\mu_\texttt{B}$, in agreement with the theoretical value of 10.334 $\mu_\texttt{B}$, and a CW temperature of $-$30.0(1) K (Figure~\ref{MErCO}B, Table~\ref{CW-parameter}).

No magnetic hysteresis loop appeared in the \emph{M}-\emph{B} measurements (Figure~\ref{MErCO}C). In contrast, clear hysteresis loops were observed previously for polycrystalline ErCrO$_3$ samples \citep{Shi2018}. For the magnetization curve at 1.8 K, a gate magnetic field of $\sim$ 650 Oe existed. When $0 \leq B \leq B_{\texttt{gate}} = 650$ Oe, the magnetization increased linearly from 0 to $\sim$ 0.45 emu g$^{-1}$. Then, it quickly flattened when $B \sim$ 0.617 T and attained $M_\texttt{meas}$ = 6.385(1) $\mu_\texttt{B}$ at 14 T (Figure~S2E).

\subsubsection{TmCrO$_3$}

We determined that $T^\texttt{Cr}_\texttt{N}$ = 125.9(1) K for the TmCrO$_3$ single crystal (Figure~\ref{MTmCO}A), which is consistent with a previous study of polycrystalline TmCrO$_3$ \citep{Wang2016-1,Yoshii2012}. Below $T^\texttt{Cr}_\texttt{N}$, the ZFC magnetization at 0 Oe was negative. The magnetization at 50 and 100 Oe was positive and increased sharply within $\sim$ 1 K. This is because Cr$^{3+}$ spins order into a canted AFM state of the $\Gamma_2$ configuration ($F_x$$C_y$$G_z$) \citep{Yoshii2012, Tamaki1977}. Upon further cooling, the magnetization increased smoothly until it reached a maximum at $T^\texttt{Cr}_\texttt{max}$ $\sim$ 74 K. After that, the magnetization reduced smoothly and attained negative values suddenly at $T_\texttt{SR}$ $\sim$ 37.2 K (for ZFC at 50 and 100 Oe) and $\sim$ 40.1 K (for FC at 100 Oe), followed by a sharp increase with positive values appearing again at a compensation temperature $T_\texttt{comp}$ $\sim$ 28.72 K. The sharp drop observed at $\sim$ 40.1 K could be ascribed to the spin reorientation of Cr$^{3+}$ ions accompanied by a 90$^\circ$ rotation of the spins, that is, from one crystallographic axis to another, probably because of the competition between anisotropic exchanges and single-ion anisotropy \citep{HFLi2016}. This feature became smooth in the polycrystalline samples \citep{Wang2016-1, Yoshii2012}. Our study revealed a reversal of the magnetization behavior. The ZFC and FC magnetization at 100 Oe nearly coincided with each other. This is different from the observations with polycrystalline TmCrO$_3$ \citep{Yoshii2012}. We observed a magnetic phase transition at $T^\texttt{Tm}_\texttt{N}$ $\sim$ 19.6 K, which probably correlates with the AFM order of Tm$^{3+}$ ions. This was not observed in polycrystalline TmCrO$_3$ samples \citep{Wang2016-1, Yoshii2012}.

\begin{table*}[!t]
\small
\caption{Calculated nearest-neighbour (NN) exchange parameters $J_1$ and $J_2$, as well as the ratio $J_2$/$J_1$, N\'{e}el temperature ($T^\texttt{MFA}_\texttt{N}$) based on the mean-field approximation, $t$-$e$ orbital overlapping degree (\emph{I}$_{t_{\texttt{2g$\downarrow$}}-e_{\texttt{g$\uparrow$}}}$), and the ordered effective moment ($M_\texttt{Cr$^{3+}$}$) of Cr$^{3+}$ ions in RECrO$_3$ orthochromates.}
\label{DFT-parameters}
\setlength{\tabcolsep}{7.0mm}{}
\renewcommand{\arraystretch}{1.05}
\begin{tabular}{lcccccc}
\hline
\hline
Parameter      & $J_1$         & $J_2$        & $J_2$/$J_1$    & $T^\texttt{MFA}_\texttt{N}$    & $I_{t_{\texttt{2g$\downarrow$}}-e_{\texttt{g$\uparrow$}}}$     & $M_\texttt{Cr$^{3+}$}$        \\ [2pt]
(unit)         & (meV)         & (meV)        &                & (K)                            & (states$^2$/eV)                                                & ($\mu_\texttt{B}$)            \\ [2pt]
\hline
EuCrO$_3$      & $-$1.420      & $-$1.390     & 0.98           & 245.7                          & $-$2.2104                                                      & 2.933                         \\
GdCrO$_3$      & $-$1.330      & $-$1.210     & 0.91           & 224.8                          & $-$2.2743                                                      & 2.931                         \\
TbCrO$_3$      & $-$1.140      & $-$0.940     & 0.82           & 186.5                          & $-$2.3571                                                      & 2.929                         \\
DyCrO$_3$      & $-$1.050      & $-$0.760     & 0.72           & 165.9                          & $-$2.4040                                                      & 2.928                         \\
YCrO$_3$       & $-$0.990      & $-$0.630     & 0.63           & 151.0                          & $-$2.4387                                                      & 2.930                         \\
HoCrO$_3$      & $-$0.940      & $-$0.430     & 0.46           & 134.3                          & $-$2.4602                                                      & 2.928                         \\
ErCrO$_3$      & $-$0.850      & $-$0.230     & 0.27           & 112.4                          & $-$2.5064                                                      & 2.927                         \\
TmCrO$_3$      & $-$0.720      & $-$0.120     & 0.17           & 90.2                           & $-$2.5396                                                      & 2.926                         \\
YbCrO$_3$      & $-$0.690      & $-$0.001     & 0.01           & 80.5                           & $-$2.5786                                                      & 2.925                         \\
LuCrO$_3$      & $-$0.760      & $-$0.046     & 0.06           & 91.3                           & $-$2.5394                                                      & 2.925                         \\
\hline
\hline
\end{tabular}
\end{table*}

We obtained $\mu_\texttt{eff-meas}$ = 8.35 $\mu_\texttt{B}$, which is almost identical to the theoretical value $\mu^\texttt{total}_\texttt{eff-theo}$ = 8.495 $\mu_\texttt{B}$, and $\Theta_\texttt{CW}$ = $-$90.6(4) K (Figure~\ref{MTmCO}B, Table~\ref{CW-parameter}).

We observed different magnetic hysteresis loops (Figure~\ref{MTmCO}C): (i) at 20 K, a parallelogram-shaped loop with $M_{\texttt{r}}$ $\sim$ 1.4 emu g$^{-1}$ and $B_{\texttt{c}}$ $\sim$ 111 Oe. (ii) At 1.8 K, a twisted loop with $M_{\texttt{r}}$ $\sim$ 0.75 emu g$^{-1}$ and $B_{\texttt{c}}$ $\sim$ 38.2 Oe. The magnetization reached $M_\texttt{meas} =$ 2.868(1) $\mu_\texttt{B}$ at 14 T (Figure~S2F, Table~\ref{CW-parameter}).

\subsubsection{YbCrO$_3$}

We determined that $T^\texttt{Cr}_\texttt{N}$ = 117.9(1) K for the YbCrO$_3$ single crystal (Figure~\ref{MYbCO}A), which is the lowest magnetic phase transition temperature of Cr$^{3+}$ sublattices among all rare-earth orthochromates. Below $T^\texttt{Cr}_\texttt{N}$, the ZFC magnetization at 0 Oe reduced sharply to negative values, whereas the magnetization at 50 and 100 Oe increased sharply and attained the maximum values at $T^\texttt{Cr}_\texttt{max}$ $\sim$ 99.3 K (for ZFC at 50 Oe), 86.3 K (for ZFC at 100 Oe), and 69.6 K (for FC at 100 Oe), followed by smooth decreases to negative values at compensation temperatures $T_\texttt{comp1}$ $\sim$ 63.7 K (for ZFC at 50 Oe), 47.5 K (for ZFC at 100 Oe), and 15.5 K (for FC at 100 Oe). The positive values of magnetization at 0, 50, and 100 (ZFC) Oe reappeared at $T_\texttt{comp2}$ $\sim$ 15.5 K (for ZFC at 0 Oe), 17.6 K (for ZFC at 50 Oe), and 19.5 K (for ZFC at 100 Oe), whereas the magnetization at FC 100 Oe still remained negative. Below $T^\texttt{Yb}_\texttt{N}$ $\sim$ 7 K, all the ZFC magnetization curves flattened \citep{Su2010-2}. Notably, a large difference exists between the curves of the ZFC and FC magnetizations at 100 Oe. The FC magnetization at 100 Oe resembles that observed in the polycrystalline samples \citep{Wang2016-2}.

The CW fitting resulted in $\mu_\texttt{eff-meas}$ = 5.63 $\mu_\texttt{B}$, which was slightly lower than the theoretical value $\mu^\texttt{total}_\texttt{eff-theo}$ = 5.964 $\mu_\texttt{B}$, and $\Theta_\texttt{CW}$ = $-$17.4(4) K (Figure~\ref{MYbCO}B, Table~\ref{CW-parameter}).

No magnetic hysteresis loop appeared in the ZFC \emph{M}-\emph{B} curves at 15 and 300 K, whereas we observed magnetic hysteresis loops with a similar shape at 1.8 K ($M_{\texttt{r}}$ $\sim$ 0.24 emu g$^{-1}$; $B_{\texttt{c}}$ $\sim$ 1200 Oe), 25 K ($M_{\texttt{r}}$ $\sim$ 0.14 emu g$^{-1}$; $B_{\texttt{c}}$ $\sim$ 1200 Oe), and 40 K ($M_{\texttt{r}}$ $\sim$ 0.25 emu g$^{-1}$; $B_{\texttt{c}}$ $\sim$ 1780 Oe) (Figure~\ref{MYbCO}C). The measured magnetization $M_\texttt{meas} =$ 0.883(1) $\mu_\texttt{B}$ at 1.8 K and 14 T, which is merely $\sim$ 17.7\% of $M^\texttt{total}_\texttt{sat-theo} =$ 5 $\mu_\texttt{B}$ (Figure~S2G, Table~\ref{CW-parameter}).

\begin{figure*} [!t]
\centering \includegraphics[width=0.82\textwidth]{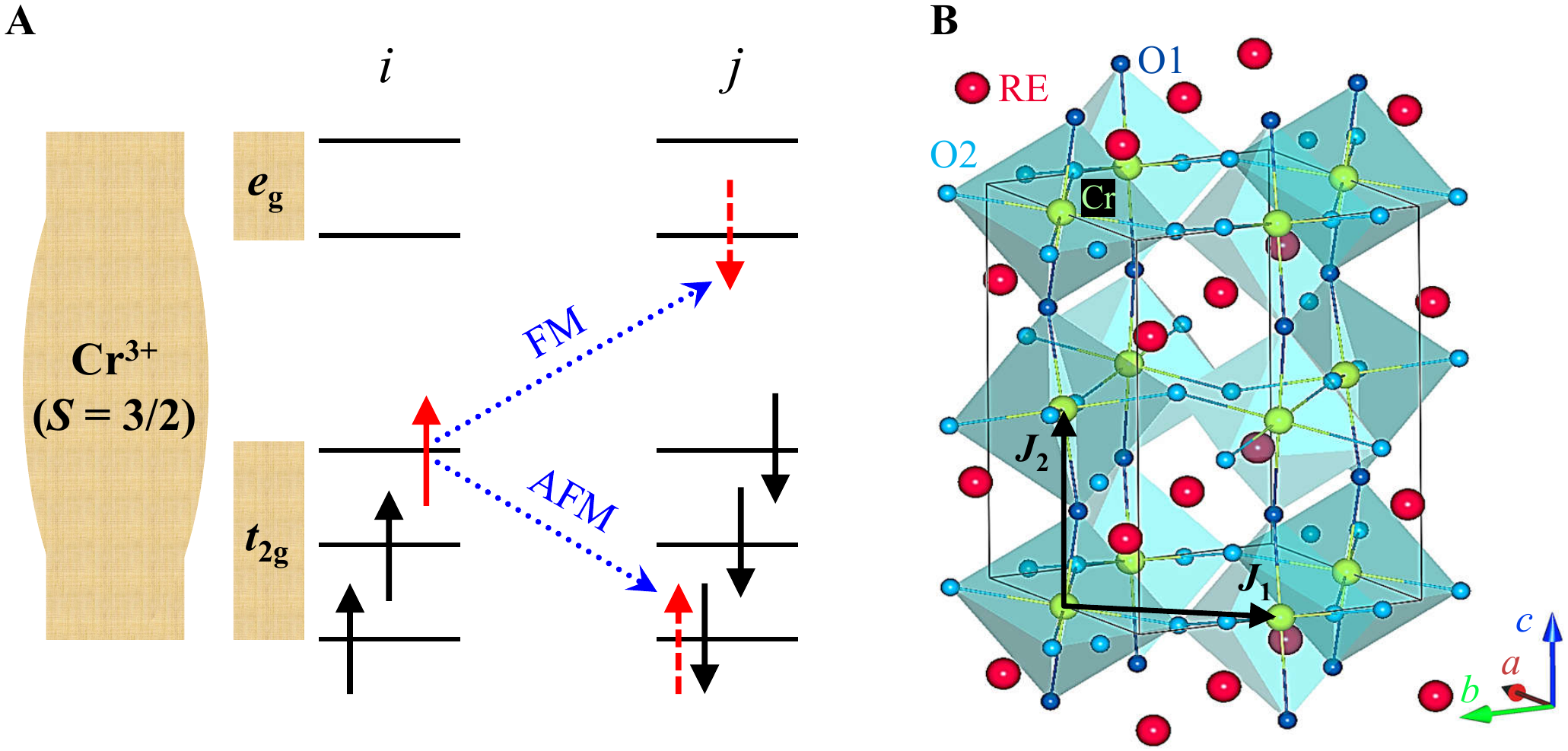}
\caption{\textbf{Process of virtual charge transfer and crystal structure} \newline
(A) Crystal field splitting of the fivefold degenerate \emph{d} orbitals of Cr$^{3+}$ ions (3$d^3$) in a cubic environment that splits the \emph{d}-level into twofold degenerate $e_\texttt{g}$ and threefold degenerate $t_{\texttt{2g}}$ levels. The arrows represent the spins of chromium. We schematically show the virtual charge transfers, leading to FM and AFM states, respectively.  \newline
(B) Refined crystal structure of RECrO$_3$ in one unit cell (solid lines) with \emph{Pbnm} space group (No. 62). The RE, Cr, O1, and O2 ions are labeled. $J_1$ and $J_2$ represent the NN spin-exchange parameters within the \emph{ab} plane and along the \emph{c} axis, respectively.
}
\label{CSCF}
\end{figure*}

\begin{figure*} [!t]
\centering \includegraphics[width=0.72\textwidth]{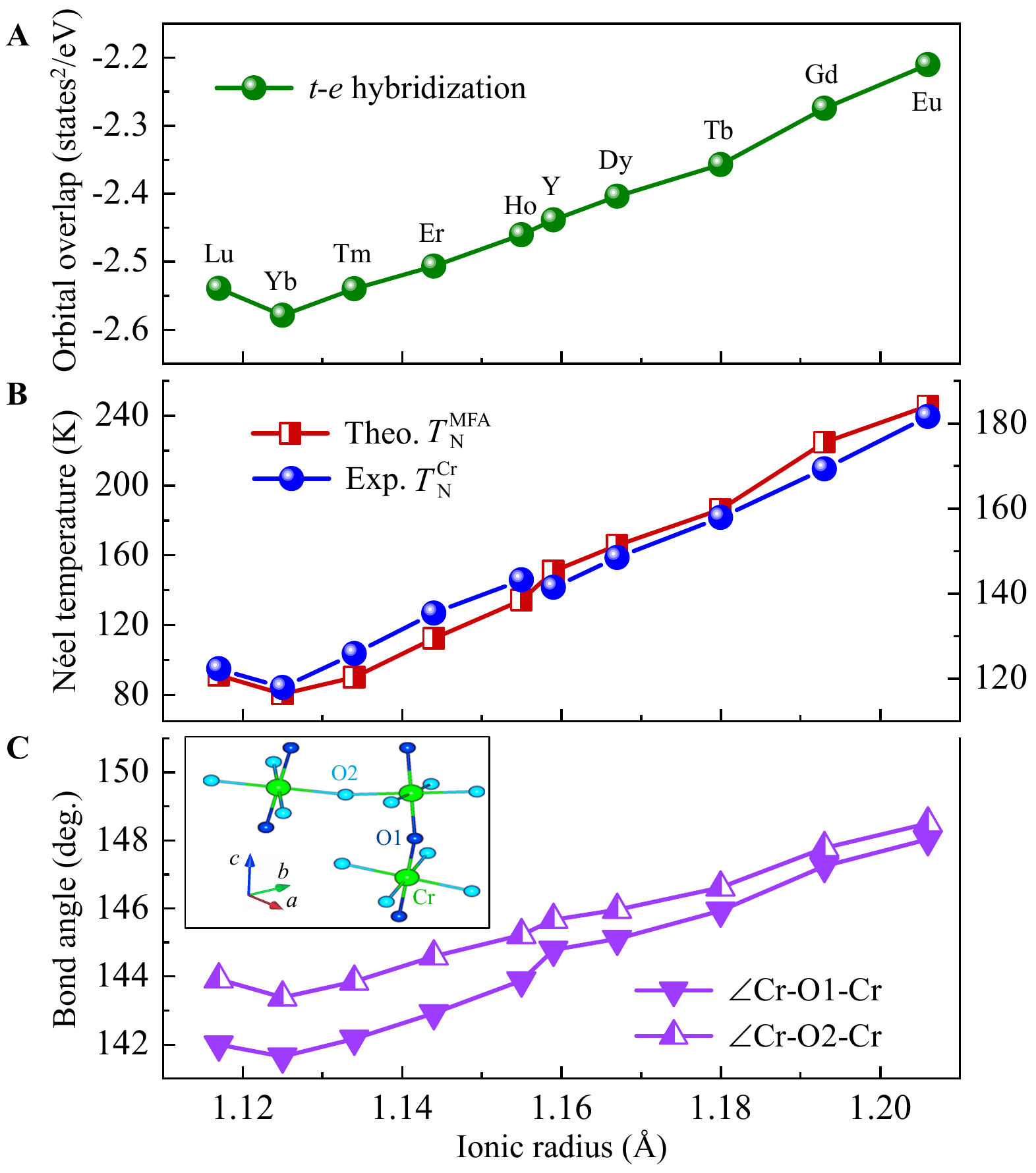}
\caption{\textbf{Coincidence between experimental and theoretical properties} \newline
(A) Calculated \emph{t}-\emph{e} orbital overlapping degree ($I_{t_{\texttt{2g{$\downarrow$}}}-e_{\texttt{g{$\uparrow$}}}}$). \newline
(B) Experimental ($T^{\texttt{Cr}}_\texttt{N}$) (right) and calculated ($T^\texttt{MFA}_\texttt{N}$) (left) AFM transition temperatures. \newline
(C) Theoretically optimized values of $\angle$Cr-O1(O2)-Cr bond angles of RECrO$_3$ (RE = Eu, Gd, Tb, Dy, Y, Ho, Er, Tm, Yb, and Lu) compounds. Inset of (C) shows the geometry of bond angles. \newline
The horizontal axis represents ionic radii of RE$^{3+}$ ions.
}
\label{TNCorr}
\end{figure*}

\subsubsection{LuCrO$_3$}

The LuCrO$_3$ single crystal exhibited magnetic behaviors similar to those observed in single-crystal TmCrO$_3$ (Figures~\ref{MLuCO}A and \ref{MTmCO}A). We determined $T^\texttt{Cr}_\texttt{N}$ = 122.3(1) K, $T^\texttt{Cr}_\texttt{max}$ $\sim$ 79 K (for ZFC and FC at 100 Oe), $T_\texttt{SR}$ $\sim$ 33 K (for ZFC at 50 Oe) and 31 K (for ZFC and FC at 100 Oe), $T_\texttt{comp2}$ $\sim$ 28 K (for all data), and $T^\texttt{Lu}_\texttt{N}$ $\sim$ 2.3 K for the LuCrO$_3$ single crystal. No clear difference was observed in the ZFC and FC magnetization at 100 Oe. Our LuCrO$_3$ single crystal demonstrated magnetic behaviors different from those of previous polycrystalline samples \citep{Duran2014}.

The CW fitting results in $\mu_\texttt{eff-meas}$ = 4.98 $\mu_\texttt{B}$, which is $\sim$ 28.1\% higher than the theoretical value $\mu^\texttt{total}_\texttt{eff-theo}$ = 3.873 $\mu_\texttt{B}$, and $\Theta_\texttt{CW}$ = $-$110.4(2) K (Figure~\ref{MLuCO}B, Table~\ref{CW-parameter}).

We observed magnetic hysteresis loops at: 1.8 K ($M_{\texttt{r}}$ $\sim$ 0.13 emu g$^{-1}$; $B_{\texttt{c}}$ $\sim$ 25 Oe), 15 K ($M_{\texttt{r}}$ $\sim$ 0.14 emu g$^{-1}$; $B_{\texttt{c}}$ $\sim$ 32 Oe), 25 K ($M_{\texttt{r}}$ $\sim$ 0.054 emu g$^{-1}$; $B_{\texttt{c}}$ $\sim$ 125 Oe), and 40 K ($M_{\texttt{r}}$ $\sim$ 0.12 emu g$^{-1}$; $B_{\texttt{c}}$ $\sim$ 28 Oe) (Figure~\ref{MLuCO}C). The measured magnetization $M_\texttt{meas} =$ 1.197(1) $\mu_\texttt{B}$ at 1.8 K and 14 T, which is merely $\sim$ 39.9\% of $M^\texttt{total}_\texttt{sat-theo} =$ 3 $\mu_\texttt{B}$ (Figure~S2H, Table~\ref{CW-parameter}).

We summarize the magnetic behaviors of single-crystal RECrO$_3$ compounds as follows: (i) TbCrO$_3$, DyCrO$_3$, and ErCrO$_3$ displayed similar temperature dependencies as did TmCrO$_3$ and LuCrO$_3$ single crystals. (ii) DyCrO$_3$ and ErCrO$_3$ did not exhibit negative magnetization, whereas the others did. (iii) Reversal magnetic behaviors (positive $\rightarrow$ negative $\rightarrow$ positive) occurred for TmCrO$_3$, YbCrO$_3$, and LuCrO$_3$ single crystals. (iv) We did not observe any indication of Eu$^{3+}$ magnetic ordering, which may require lower temperatures. (v) Obvious magnetic hysteresis loops were observed for RECrO$_3$, except for ErCrO$_3$. (vi) The measured magnetization at 1.8 K and high magnetic fields plateaued for RECrO$_3$ (RE = Tb, Dy, Ho, Er, and Lu) single crystals. (vii) The measured effective PM moments of EuCrO$_3$ and LuCrO$_3$ were not consistent with the theoretical values. (viii) The applied magnetic field of 14 T was far less to saturate RECrO$_3$ at 1.8 K. (ix) Only HoCrO$_3$ demonstrated a positive CW temperature. Finally, we observed the complex and coupled magnetic phase transitions of RE$^{3+}$ (except for Eu$^{3+}$) and Cr$^{3+}$ ions.

We also summarize the magnetic structures of RE$^{3+}$ and Cr$^{3+}$ ions in RECrO$_3$ (RE = Gd, Tb, Dy, Ho, Er, Tm, Yb, Lu) compounds in the literature (Table~\ref{Magentic-structures}). The magnetic SR transition of Cr$^{3+}$ ions has been reported for GdCrO$_3$ \citep{Cooke1974} and ErCrO$_3$ \citep{Hornreich1978, Shamir1981} compounds. The detailed magnetic structures of Tm$^{3+}$ and Yb$^{3+}$ ions in TmCrO$_3$ \citep{Shamir1981} and YbCrO$_3$ \citep{Shamir1981} compounds remain controversial. A $F_x$ magnetic component of Yb$^{3+}$ ions had to be included for a satisfactory fit of the neutron powder diffraction data of YbCrO$_3$ in the temperature range of 1.5--120 K \citep{Deepak2021}. Unraveling the nature of the magnetic phase transitions necessitates neutron scattering studies on single-crystal samples with modern techniques \citep{Li2008-1}.

\section{Discussion}

Superexchange interactions between the neighboring spins of transition metals can be realized through VCT via intermediate O$^{2-}$ ions. During this process, the tilting of the oxygen octahedron corresponds to the change in the metal-oxygen-metal bond angles and may lead to $t_\texttt{2g}$ and $e_\texttt{g}$ orbital overlapping. This facilitates the hopping of $t_\texttt{{2g}$\uparrow$}$ electrons via the bridge of O$^{2-}$ ions to occupy the empty $e_\texttt{g}$ band and the interaction with filled $t_\texttt{{2g}$\downarrow$}$ electrons at the same site, that is, the intersite $t$-$e$ orbital hybridization \citep{Zhou2010, Zhou2008}. In the framework of $t$-$e$ hybridization, the superexchange parameter \emph{J} consists of the following two parts \citep{Zhou2010}:
\begin{eqnarray}
J = J^{\pi} - J^{\sigma}_\texttt{hb},
\label{JFMAFM}
\end{eqnarray}
where $J^{\pi}$ denotes AFM coupling via the VCT of $t_\texttt{2g}^3$-O-$t_\texttt{2g}^3$, and $J^{\sigma}_{\texttt{hb}}$ represents FM coupling via the VCT of $t_\texttt{2g}^3$-O-$e_\texttt{g}^2$. Both processes are schematically depicted in Figure~\ref{CSCF}A for Cr$^{3+}$ ions. For example, for a half-filled transition metal like Fe$^{3+}$ ($t_\texttt{2g}^3e_\texttt{g}^2$), the effect of $t$-$e$ hybridization on superexchange interactions may not be evident because the electron hoppings of Fe$^{3+}$ ($t_\texttt{2g$\uparrow$}^3$)-O$^{2-}$-Fe$^{3+}$ ($t_\texttt{2g$\downarrow$}^3$) and Fe$^{3+}$ ($t_\texttt{2g$\uparrow$}^3$)-O$^{2-}$-Fe$^{3+}$ ($e_\texttt{g$\downarrow$}^2$) themselves are AFM couplings already \citep{Zhou2010, Zhou2008}. In contrast, for less than half-filled 3\emph{d} electrons like Cr$^{3+}$ ions ($t_\texttt{2g}^3e_\texttt{g}^0$) in RECrO$_3$ compounds, the $t$-$e$ orbital hybridization favors the VCT of Cr$^{3+}$ ($t_\texttt{2g$\uparrow$}^3$)-O$^{2-}$-Cr$^{3+}$ ($e_\texttt{g}^0$) \citep{Zhou2010,Qian2014,Siddiqu2021}. Electron hopping of $t_\texttt{2g}^3$-O-$e_\texttt{g}^0$ can increase the FM coupling component $J^{\sigma}_{\texttt{hb}}$. When RE$^{3+}$ ions change from La to Lu in RECrO$_3$, the competition between the AFM ($J^{\pi}$) and FM ($J^{\sigma}_{\texttt{hb}}$) components would probably result in a variation in $T^{\texttt{Cr}}_\texttt{N}$.

To quantitatively describe the Cr$^{3+}$-O$^{2-}$-Cr$^{3+}$ superexchange interactions as well as the $t_{\texttt{2g}}$-$e_{\texttt{g}}$ (\emph{t}-\emph{e}) orbital overlapping degree as a function of the ionic radii of the RE$^{3+}$ ions, we calculated the exchange parameters and electronic structures of RECrO$_3$ (RE = Y, Eu, Gd, Tb, Dy, Ho, Er, Tm, Yb, and Lu) compounds. We mainly considered the NN exchange parameters of the Cr$^{3+}$ sublattices within the crystallographic \emph{ab} plane ($J_1$) and along the \emph{c} axis ($J_2$) (Figure~\ref{CSCF}B). We extracted the values of $J_1$ and $J_2$ using the so-called energy mapping method with four types of magnetic structures (\emph{A}-, \emph{C}-, \emph{G}-AFM, and FM) (Figure~S3). We obtained the total energy of each magnetic structure using DFT calculations, projecting each collinear spin state onto the following spin Hamiltonian of a Heisenberg model
\begin{eqnarray}
H = -\Sigma{J_{ij}S_i{\cdot}S_j},
\label{Heisenberg}
\end{eqnarray}
where \emph{J} $>$ 0 represents FM interactions, and \emph{J} $<$ 0 denotes AFM couplings. Using equation (\ref{Heisenberg}), we can solve for $J_1$ and $J_2$ as \citep{Fujioka2008, Bernal2021}
\begin{eqnarray}
J_1 =& (E_{G}+E_{C} - E_{A} - E_{F})/8S^2,& {\texttt{and}}          \\
J_2 =& (E_{G} - E_{C}+E_{A} - E_{F})/4S^2.&
\label{J12}
\end{eqnarray}
Thus, we can calculate the N\'{e}el temperatures of the RECrO$_3$ compounds using the mean-field approximation (MFA) \citep{Fujioka2008}, that is,
\begin{eqnarray}
T_\texttt{N} = \frac{2S(S+1)}{3k_B}(-4J_1 - 2J_2).
\label{TN}
\end{eqnarray}

To depict the hybridization degree, we calculated the overlap of $t_\texttt{2g$\downarrow$}$ and $e_\texttt{g$\uparrow$}$ (Table~\ref{DFT-parameters}) using their DOS (Figure~S4) product over the corresponding energy region.
\begin{eqnarray}
I_{t_{\texttt{2g}\downarrow}-e_{\texttt{g}\uparrow}} = \int_{E_\texttt{F}}^{E_{\texttt{c}}}D_{t_{\texttt{2g}\downarrow}}(E)D_{e_\texttt{g}\uparrow}(E)dE,
\label{overlap}
\end{eqnarray}
where $D_{t_{\texttt{2g}\downarrow}}(E)$ and $D_{e_{\texttt{g}\uparrow}}(E)$ are the DOSs for the $t_{\texttt{2g}\downarrow}$ and unoccupied $e_{\texttt{g}\uparrow}$ states of Cr$^{3+}$ ions, respectively. $E_\texttt{F}$ and $E_\texttt{c}$ represent the Fermi level and cutoff energy of the $t_\texttt{2g$\downarrow$}$-$e_\texttt{g$\uparrow$}$ hybridization, respectively. With Equation (\ref{overlap}), we could quantitatively describe the \emph{t}-\emph{e} hybridization in RECrO$_3$ orthochromates (Table~\ref{DFT-parameters}).

Different $U_\texttt{eff}$-value settings have a relatively minor effect on the \emph{t}-\emph{e} hybridization; therefore, our calculations are based on the on-site Coulomb interactions between Cr$^{3+}$ ions using Hubbard $U_\texttt{eff}$ = 3.3 eV derived under the framework of a linear-response ansatz, which is in good agreement with the measured band gaps \citep{Singh2018}. Meanwhile, the electron dispersions are not largely influenced by fixing $U_\texttt{eff}$, which assures the calculation of reliable \emph{t}-\emph{e} orbital hybridizations of RECrO$_3$ orthochromates by considering only the influence of RE$^{3+}$ ions.

Based on the foregoing discussion, we first optimized the structural parameters of RECrO$_3$ orthochromates (Tables S1 and S2). Our calculations also indicate that the \emph{G}-type AFM is the most stable magnetic structure for all the RECrO$_3$ orthochromates (Figure~S3 and Table S3). Subsequently, we extracted the optimized values of $J_1$, $J_2$, N\'{e}el temperature ($T_\texttt{N}^\texttt{MFA}$), and the $t_\texttt{2g}$-$e_\texttt{g}$ orbital overlapping degree ($I_{t_\texttt{2g}-e_\texttt{g}}$) (Table~\ref{DFT-parameters}).

When RE$^{3+}$ ions varies from Eu to Lu, the changes of the calculated $I_{t_\texttt{2g}-e_\texttt{g}}$ (Figure~\ref{TNCorr}A), $T_\texttt{N}^\texttt{MFA}$ (Figure~\ref{TNCorr}B), and $\angle$Cr-O1-Cr and $\angle$Cr-O2-Cr (Figure~\ref{TNCorr}C) demonstrated a similar trend, indicating a strong correlation between them and a clear effect of the RE$^{3+}$ radii on the superexchange interactions. The calculated $T_\texttt{N}^\texttt{MFA}$ coincided with the experimental $T^{\texttt{Cr}}_\texttt{N}$ values (Table~\ref{CW-parameter}, Figure~\ref{TNCorr}B). The average value ($\langle$$\angle$Cr-O-Cr$\rangle$) of the $\angle$Cr-O1(2)-Cr bond angles changed from 148.34$^\circ$ (EuCrO$_3$) to 142.81$^\circ$ (YbCrO$_3$) (Tables~S1 and S2), leading to the \emph{t}-\emph{e} hybridization ($I_{t_\texttt{2g}-e_\texttt{g}}$) changing from -2.2104 (EuCrO$_3$) to -2.5786 states$^2$/eV (YbCrO$_3$) and the corresponding $T_\texttt{N}^\texttt{MFA}$ decreased from 245.7 (EuCrO$_3$) to 80.5 K (YbCrO$_3$) (Table~\ref{DFT-parameters}). Therefore, the decrease in bond angles of $\angle$Cr-O1(2)-Cr facilitated \emph{t}-\emph{e} hybridization by enhancing the FM component ($J^{\sigma}_{\texttt{hb}}$) within the entire superexchange interaction. Both our experimental and theoretical studies produced the minimum AFM transition temperature for YbCrO$_3$ in the system (Figure~\ref{TNCorr}B), which is inconsistent with a previous study on polycrystal RECrO$_3$ samples, where the minimum $T_\texttt{N}$, as well as $\langle$$\angle$Cr-O-Cr$\rangle$, occurred to LuCrO$_3$ \citep{Zhou2008}.

Our DFT calculations demonstrate that the magnetic anisotropy of Cr$^{3+}$ sublattices can be tuned by RE$^{3+}$ ions. The ratio of $J_2$/$J_1$ is a good parameter that inversely expresses anisotropy. It indeed reduces from $J_2$/$J_1$ $\sim$ 0.98 (EuCrO$_3$) to $\sim$ 0.01 (YbCrO$_3$) (Table~\ref{DFT-parameters}) and strongly correlates with the values of the $\angle$Cr-O1(2)-Cr bond angles. The $\angle$Cr-O1-Cr angle was outside the \emph{ab} plane, and its value was smaller than that of $\angle$Cr-O2-Cr within the \emph{ab} plane (Inset, Figure~\ref{TNCorr}C). Therefore, we infer that the VCT of $t_\texttt{2g$\uparrow$}^3$-O$^{2-}$-$t_\texttt{2g$\downarrow$}^3$ produces the major AFM spin interactions within the \emph{ab} plane, whereas the VCT of $t_\texttt{2g$\uparrow$}^3$-O$^{2-}$-$e_\texttt{g}^0$ can generate strong FM couplings along the \emph{c} axis. This is in good agreement with the previously proposed magnetic model \citep{Ziel1969}. The difference between the $\angle$Cr-O1(2)-Cr values increases when RE$^{3+}$ varies from Eu to Lu, indicating that the competition between in-plane AFM interactions and out-of-plane FM couplings becomes increasingly strong with an enhancement of FM interactions. The largest difference between $\angle$Cr-O1-Cr and $\angle$Cr-O2-Cr occurred for YbCrO$_3$, leading to the smallest $J_2$/$J_1$ ratio ($\sim$ 0.01) and, thus, the largest magnetic anisotropy.

The FM coupling $J^{\sigma}_\texttt{hb}$ directly acts on the overall Cr-O-Cr superexchange. This is induced by the $t_\textrm{2g}$-$e_\textrm{g}$ orbital hybridization and could compete with AFM coupling, leading to an overall exchange $J = J^{\pi} - J^{\sigma}_\texttt{hb}$ \citep{Zhou2010}. The state of Cr$^{3+}$, that is, the 3$d^3$ configuration with half-filled $t_\textrm{2g}$ orbital and empty $e_\textrm{g}$ orbital, makes it a special ion compared with other elements such as V, Fe, and Mn \citep{Streltsov2008}. By chemical engineering \citep{Yaresko2008} or applying high pressure \citep{Fita2021}, one may strengthen $t_\textrm{2g}$-$e_\textrm{g}$ hopping but simultaneously weaken $t_\textrm{2g}$-$t_\textrm{2g}$ hopping. Eventually, this could result in an overall FM component superexchange between neighbouring Cr$^{3+}$ cations. The weak ferromagnetism results from competition between the Heisenberg exchange and DM interactions in orthochromates. This is one case of an AFM structure with a small canting, leading to a noncollinear spin configuration and thus a net FM component. Unravelling the competing degree of different spin interactions necessitates inelastic neutron scattering studies on single-crystal orthochromates.

\section{Conclusion}

We have successfully grown a series of RECrO$_3$ (RE = Y, Eu--Lu) single crystals with a laser-diode FZ furnace. The grown crystals are centimeter (gram)-scale with a good quality. We performed magnetization measurements as functions of temperature and applied magnetic field, providing more reliable and intrinsic magnetic properties. We investigated theoretically the $t$-$e$ hybridization by quantitatively calculating the $t_{\texttt{2g}}$ and $e_{\texttt{g}}$ orbital overlapping degree ($I_{t_{\texttt{2g$\downarrow$}}-e_{\texttt{g$\uparrow$}}}$) based on DOS calculations. As RE$^{3+}$ ions change from Eu to Lu, the calculated AFM transition temperatures demonstrate a similar trend with those determined experimentally. The changes in the $\angle$Cr-O1(2)-Cr bond angles strongly influence the weight factor of FM couplings within the entire superexchanges interactions by (dis)favoring the VCT of $t_\texttt{2g$\uparrow$}^3$-O$^{2-}$-$e_\texttt{g}^0$. This may be the origin of the weak ferromagnetism appearing within the main AFM matrix of RECrO$_3$. The difference between $\angle$Cr-O1(2)-Cr bond angles results in a magnetic anisotropy between within the \emph{ab} plane and along the \emph{c} axis. The change of \emph{t}-\emph{e} hybridization coincides well with that of $\angle$Cr-O1(2)-Cr and that of $T^\texttt{MFA}_\texttt{N}$. Our research sheds light on the origin of the intriguing magnetism in RECrO$_3$ system.

\section*{Limitations of the study}

The single crystal growth of orthochromates is extremely difficult. First, traditional FZ furnaces equipped with four IR-heating halogen lamps and four ellipsoidal mirrors (such as Model FZ-T-10000-H-VI-VPO-PC from Crystal Systems Inc.) cannot even access the melting temperature of orthochromates so that the seed and feed rods could not be melted, thus the crystal growth could not be performed. Second, the evaporation of Cr-based oxides is very heavy like a thick haze. Third, the process of crystal growth is time-consuming and labor-intensive, and one needs to optimize various growth parameters. Presently, we are unable to measure magnetic properties as a function of the crystallographic orientation because of technique difficulties, which is left for future work.

\section*{SUPPLEMENTAL INFORMATION}

Supplemental information can be found online at xxx.

\section*{ACKNOWLEDGMENTS}

The work at City University of Hong Kong was supported by grants from the Research Grants Council of the Hong Kong SAR (Project Nos. 11305618 and 11306219) and City University of Hong Kong (SRG-Fd Project No. 7005496 and SIRG Project No. 7020017).
The work at State Key Laboratory of High Performance Ceramics and Superfine Microstructure, Shanghai Institute of Ceramics, Chinese Academy of Sciences, was supported by the Technology Commission of Shanghai Municipality (19DZ1100703 and 19511107600), the Research Program of Chinese Academy of Sciences (YJ- 267 KYYQ20180025).
The work at University of Macau was supported by the opening project of State Key Laboratory of High Performance Ceramics and Superfine Microstructure (Grant No. SKL201907SIC), Science and Technology Development Fund, Macao SAR (File Nos. 0051/2019/AFJ and 0090/2021/A2), Guangdong Basic and Applied Basic Research Foundation (Guangdong-Dongguan Joint Fund No. 2020B1515120025), University of Macau (MYRG2020-00278-IAPME and EF030/IAPME-LHF/2021/GDSTIC), and Guangdong-Hong Kong-Macao Joint Laboratory for Neutron Scattering Science and Technology (Grant No. 2019B121205003).

\section*{AUTHOR CONTRIBUTIONS}

Y.H.Z., J.C.X., and S.W. contributed equally.

Y.H.Z., J.C.X., S.W., Y.Z., L.W., H.W., J.H.F., C.Y.W., T.W., Y.S., and J.D.Y. grew the single crystals.
Y.H.Z., S.W., Y.Z., L.W., and H.W. performed the Laue experiments.
Y.H.Z., J.C.X., S.W., and K.T.S. performed magnetization measurements.
Y.H.Z., Y.W.Y., Y.L.Z., H.W.K., and R.Q.Z. carried out the theoretical calculations.
All authors discussed and analyzed the results.
Y.H.Z. and H.F.L. wrote the main manuscript text.
All authors reviewed the paper.
R.Q.Z. and H.F.L. conceived and directed the project.

\section*{DECLARATION OF INTERESTS}

The authors declare the following competing financial interest(s): Y. H. Zhu, S. Wu, and H.-F. Li have a 2021 China Invention Patent (CN110904497B) through University of Macau based on this work: A method of centimeter-sized single crystal growth of chromate compounds and related storage device.

\bibliographystyle{model5-names}
\biboptions{authoryear}
\bibliography{RCOSC2}

\begin{thebibliography}{82}
\expandafter\ifx\csname natexlab\endcsname\relax\def\natexlab#1{#1}\fi
\providecommand{\url}[1]{\texttt{#1}}
\providecommand{\href}[2]{#2}
\providecommand{\path}[1]{#1}
\providecommand{\DOIprefix}{doi:}
\providecommand{\ArXivprefix}{arXiv:}
\providecommand{\URLprefix}{URL: }
\providecommand{\Pubmedprefix}{pmid:}
\providecommand{\doi}[1]{\href{http://dx.doi.org/#1}{\path{#1}}}
\providecommand{\Pubmed}[1]{\href{pmid:#1}{\path{#1}}}
\providecommand{\bibinfo}[2]{#2}
\ifx\xfnm\relax \def\xfnm[#1]{\unskip,\space#1}\fi
\bibitem[{Bernal et~al.(2021)Bernal, Lundvall, Kumar, Hansen, Wragg,
  Fjellv{\aa}g \& L{\o}vvik}]{Bernal2021}
\bibinfo{author}{Bernal, F. L.~M.}, \bibinfo{author}{Lundvall, F.},
  \bibinfo{author}{Kumar, S.}, \bibinfo{author}{Hansen, P.-A.~S.},
  \bibinfo{author}{Wragg, D.~S.}, \bibinfo{author}{Fjellv{\aa}g, H.}, \&
  \bibinfo{author}{L{\o}vvik, O.~M.} (\bibinfo{year}{2021}).
\newblock \bibinfo{title}{Jahn-teller active fluoroperovskites {ACrF}$_{3}$ ({A
  = Na}$^{+}$, {K}$^{+}$): Magnetic and thermo-optical properties}.
\newblock {\it \bibinfo{journal}{Phys. Rev. Mater.}\/},  {\it
  \bibinfo{volume}{5}\/}, \bibinfo{pages}{064420}.
\bibitem[{Bertaut(1963)}]{Bertaut1963}
\bibinfo{author}{Bertaut, E.~F.} (\bibinfo{year}{1963}).
\newblock \bibinfo{title}{Magnetism iii, edited by g. rado and h. suhl}.
\newblock (p. \bibinfo{pages}{149}).
\newblock \bibinfo{publisher}{Academic Press, New York}.
\bibitem[{Besbes et~al.(2019)Besbes, Nikolaev, Meskini \&
  Solovyev}]{Besbes2019}
\bibinfo{author}{Besbes, O.}, \bibinfo{author}{Nikolaev, S.},
  \bibinfo{author}{Meskini, N.}, \& \bibinfo{author}{Solovyev, I.}
  (\bibinfo{year}{2019}).
\newblock \bibinfo{title}{Microscopic origin of ferromagnetism in the
  trihalides {CrCl}$_{3}$ and {CrI}$_{3}$}.
\newblock {\it \bibinfo{journal}{Phys. Rev. B}\/},  {\it
  \bibinfo{volume}{99}\/}, \bibinfo{pages}{104432}.
\bibitem[{Chatterji et~al.(2017)Chatterji, Demmel, Jalarvo, Podlesnyak, Kumar,
  Xiao \& Br{\"u}ckel}]{Chatterji2017}
\bibinfo{author}{Chatterji, T.}, \bibinfo{author}{Demmel, F.},
  \bibinfo{author}{Jalarvo, N.}, \bibinfo{author}{Podlesnyak, A.},
  \bibinfo{author}{Kumar, C.}, \bibinfo{author}{Xiao, Y.}, \&
  \bibinfo{author}{Br{\"u}ckel, T.} (\bibinfo{year}{2017}).
\newblock \bibinfo{title}{Quasielastic and low-energy inelastic neutron
  scattering study of {HoCrO}$_3$ by high resolution time-of-flight neutron
  spectroscopy}.
\newblock {\it \bibinfo{journal}{J. Phys.: Condens. Matter}\/},  {\it
  \bibinfo{volume}{29}\/}, \bibinfo{pages}{475802}.
\bibitem[{Cheng(2017)}]{CHENG20171039}
\bibinfo{author}{Cheng, H.-M.} (\bibinfo{year}{2017}).
\newblock \bibinfo{title}{Metre-size single-crystal graphene becomes a
  reality}.
\newblock {\it \bibinfo{journal}{Science Bulletin}\/},  {\it
  \bibinfo{volume}{62}\/}, \bibinfo{pages}{1039--1040}.
\bibitem[{Cococcioni \& de~Gironcoli(2005)}]{Cococcioni2005}
\bibinfo{author}{Cococcioni, M.}, \& \bibinfo{author}{de~Gironcoli, S.}
  (\bibinfo{year}{2005}).
\newblock \bibinfo{title}{Linear response approach to the calculation of the
  effective interaction parameters in the {LDA} + {U} method}.
\newblock {\it \bibinfo{journal}{Phys. Rev. B}\/},  {\it
  \bibinfo{volume}{71}\/}, \bibinfo{pages}{035105}.
\bibitem[{Coffey et~al.(1991)Coffey, Rice \& Zhang}]{Coffey1992}
\bibinfo{author}{Coffey, D.}, \bibinfo{author}{Rice, T.~M.}, \&
  \bibinfo{author}{Zhang, F.~C.} (\bibinfo{year}{1991}).
\newblock \bibinfo{title}{{Dzyaloshinskii-Moriya} interaction in the cuprates}.
\newblock {\it \bibinfo{journal}{Phys. Rev. B}\/},  {\it
  \bibinfo{volume}{44}\/}, \bibinfo{pages}{10112--10116}.
\bibitem[{Cooke et~al.(1974)Cooke, Martin \& Wells}]{Cooke1974}
\bibinfo{author}{Cooke, A.~H.}, \bibinfo{author}{Martin, D.~M.}, \&
  \bibinfo{author}{Wells, M.~R.} (\bibinfo{year}{1974}).
\newblock \bibinfo{title}{Magnetic interactions in gadolinium orthochromite,
  {GdCrO}$_3$}.
\newblock {\it \bibinfo{journal}{J. Phys. C: Solid State Phys.}\/},  {\it
  \bibinfo{volume}{7}\/}, \bibinfo{pages}{3133--3144}.
\bibitem[{Deepak et~al.(2021)Deepak, Kumar \& Yusuf}]{Deepak2021}
\bibinfo{author}{Deepak}, \bibinfo{author}{Kumar, A.}, \&
  \bibinfo{author}{Yusuf, S.~M.} (\bibinfo{year}{2021}).
\newblock \bibinfo{title}{Intertwined magnetization and exchange bias reversals
  across compensation temperature in {YbCrO}$_{3}$ compound}.
\newblock {\it \bibinfo{journal}{Phys. Rev. Mater.}\/},  {\it
  \bibinfo{volume}{5}\/}, \bibinfo{pages}{124402}.
\bibitem[{Dmitrienko et~al.(2014)Dmitrienko, Ovchinnikova, Collins, Nisbet,
  Beutier, Kvashnin, Mazurenko, Lichtenstein \& Katsnelson}]{Dmitrienko2014}
\bibinfo{author}{Dmitrienko, V.~E.}, \bibinfo{author}{Ovchinnikova, E.~N.},
  \bibinfo{author}{Collins, S.~P.}, \bibinfo{author}{Nisbet, G.},
  \bibinfo{author}{Beutier, G.}, \bibinfo{author}{Kvashnin, Y.~O.},
  \bibinfo{author}{Mazurenko, V.~V.}, \bibinfo{author}{Lichtenstein, A.~I.}, \&
  \bibinfo{author}{Katsnelson, M.~I.} (\bibinfo{year}{2014}).
\newblock \bibinfo{title}{Measuring the {Dzyaloshinskii-Moriya} interaction in
  a weak ferromagnet}.
\newblock {\it \bibinfo{journal}{Nat. Phys.}\/},  {\it \bibinfo{volume}{10}\/},
  \bibinfo{pages}{202--206}.
\bibitem[{Dudarev et~al.(1998)Dudarev, Botton, Savrasov, Humphreys \&
  Sutton}]{Dudarev1998}
\bibinfo{author}{Dudarev, S.~L.}, \bibinfo{author}{Botton, G.~A.},
  \bibinfo{author}{Savrasov, S.~Y.}, \bibinfo{author}{Humphreys, C.~J.}, \&
  \bibinfo{author}{Sutton, A.~P.} (\bibinfo{year}{1998}).
\newblock \bibinfo{title}{Electron-energy-loss spectra and the structural
  stability of nickel oxide: An lsda+u study}.
\newblock {\it \bibinfo{journal}{Phys. Rev. B}\/},  {\it
  \bibinfo{volume}{57}\/}, \bibinfo{pages}{1505--1509}.
\bibitem[{Dur{\'a}n et~al.(2014)Dur{\'a}n, Mor{\'a}n, Alario-Franco, Ostos
  et~al.}]{Duran2014}
\bibinfo{author}{Dur{\'a}n, A.}, \bibinfo{author}{Mor{\'a}n, E.},
  \bibinfo{author}{Alario-Franco, M.}, \bibinfo{author}{Ostos, C.} et~al.
  (\bibinfo{year}{2014}).
\newblock \bibinfo{title}{Biferroic {LuCrO}$_{3}$: structural characterization,
  magnetic and dielectric properties}.
\newblock {\it \bibinfo{journal}{Mater. Chem. Phys.}\/},  {\it
  \bibinfo{volume}{143}\/}, \bibinfo{pages}{1222--1227}.
\bibitem[{Eibsch{\"u}tz et~al.(1970)Eibsch{\"u}tz, Holmes, Maita \&
  Van~Uitert}]{Eibschutz1970}
\bibinfo{author}{Eibsch{\"u}tz, M.}, \bibinfo{author}{Holmes, L.},
  \bibinfo{author}{Maita, J.~P.}, \& \bibinfo{author}{Van~Uitert, L.~G.}
  (\bibinfo{year}{1970}).
\newblock \bibinfo{title}{Low temperature magnetic phase transition in
  {ErCrO}$_{3}$}.
\newblock {\it \bibinfo{journal}{Solid State Commun.}\/},  {\it
  \bibinfo{volume}{8}\/}, \bibinfo{pages}{1815--1817}.
\bibitem[{{El Amrani} et~al.(2014){El Amrani}, Zaghrioui, {Ta Phuoc}, Gervais
  \& Massa}]{Amrani2014}
\bibinfo{author}{{El Amrani}, M.}, \bibinfo{author}{Zaghrioui, M.},
  \bibinfo{author}{{Ta Phuoc}, V.}, \bibinfo{author}{Gervais, F.}, \&
  \bibinfo{author}{Massa, N.~E.} (\bibinfo{year}{2014}).
\newblock \bibinfo{title}{Local symmetry breaking and spin–phonon coupling in
  {SmCrO}$_3$ orthochromite}.
\newblock {\it \bibinfo{journal}{J. Magn. Magn. Mater.}\/},  {\it
  \bibinfo{volume}{361}\/}, \bibinfo{pages}{1--6}.
\bibitem[{Fiebig et~al.(2016)Fiebig, Lottermoser, Meier \&
  Trassin}]{Fiebig2016}
\bibinfo{author}{Fiebig, M.}, \bibinfo{author}{Lottermoser, T.},
  \bibinfo{author}{Meier, D.}, \& \bibinfo{author}{Trassin, M.}
  (\bibinfo{year}{2016}).
\newblock \bibinfo{title}{The evolution of multiferroics}.
\newblock {\it \bibinfo{journal}{Nat. Rev. Mater.}\/},  {\it
  \bibinfo{volume}{1}\/}, \bibinfo{pages}{16046}.
\bibitem[{Fita et~al.(2021)Fita, Puzniak \& Wisniewski}]{Fita2021}
\bibinfo{author}{Fita, I.}, \bibinfo{author}{Puzniak, R.}, \&
  \bibinfo{author}{Wisniewski, A.} (\bibinfo{year}{2021}).
\newblock \bibinfo{title}{Pressure-tuned spin switching in compensated
  {GdCrO}$_{3}$ ferrimagnet}.
\newblock {\it \bibinfo{journal}{Phys. Rev. B}\/},  {\it
  \bibinfo{volume}{103}\/}, \bibinfo{pages}{054423}.
\bibitem[{Franchini et~al.(2007{\natexlab{a}})Franchini, Podloucky, Paier,
  Marsman \& Kresse}]{Anisimov1997}
\bibinfo{author}{Franchini, C.}, \bibinfo{author}{Podloucky, R.},
  \bibinfo{author}{Paier, J.}, \bibinfo{author}{Marsman, M.}, \&
  \bibinfo{author}{Kresse, G.} (\bibinfo{year}{2007}{\natexlab{a}}).
\newblock \bibinfo{title}{Ground-state properties of multivalent manganese
  oxides: Density functional and hybrid density functional calculations}.
\newblock {\it \bibinfo{journal}{Phys. Rev. B}\/},  {\it
  \bibinfo{volume}{75}\/}, \bibinfo{pages}{195128}.
\bibitem[{Franchini et~al.(2007{\natexlab{b}})Franchini, Podloucky, Paier,
  Marsman \& Kresse}]{Franchini2007}
\bibinfo{author}{Franchini, C.}, \bibinfo{author}{Podloucky, R.},
  \bibinfo{author}{Paier, J.}, \bibinfo{author}{Marsman, M.}, \&
  \bibinfo{author}{Kresse, G.} (\bibinfo{year}{2007}{\natexlab{b}}).
\newblock \bibinfo{title}{Ground-state properties of multivalent manganese
  oxides: Density functional and hybrid density functional calculations}.
\newblock {\it \bibinfo{journal}{Phys. Rev. B}\/},  {\it
  \bibinfo{volume}{75}\/}, \bibinfo{pages}{195128}.
\bibitem[{Fujioka et~al.(2008)Fujioka, Frantti \& Nieminen}]{Fujioka2008}
\bibinfo{author}{Fujioka, Y.}, \bibinfo{author}{Frantti, J.}, \&
  \bibinfo{author}{Nieminen, R.~M.} (\bibinfo{year}{2008}).
\newblock \bibinfo{title}{Electronic energy band structure of the double
  perovskite {Ba}$_{2}${MnWO}$_{6}$}.
\newblock {\it \bibinfo{journal}{J. Phys. Chem. B}\/},  {\it
  \bibinfo{volume}{112}\/}, \bibinfo{pages}{6742--6746}.
\bibitem[{Gordon et~al.(1976)Gordon, Hornreich, Shtrikman \&
  Wanklyn}]{Gordon1976}
\bibinfo{author}{Gordon, J.~D.}, \bibinfo{author}{Hornreich, R.~M.},
  \bibinfo{author}{Shtrikman, S.}, \& \bibinfo{author}{Wanklyn, B.~M.}
  (\bibinfo{year}{1976}).
\newblock \bibinfo{title}{Magnetization studies in the rare-earth
  orthochromites. v. {TbCrO}$_{3}$ and {PrCrO}$_{3}$}.
\newblock {\it \bibinfo{journal}{Phys. Rev. B}\/},  {\it
  \bibinfo{volume}{13}\/}, \bibinfo{pages}{3012--3017}.
\bibitem[{Hornreich(1978)}]{Hornreich1978}
\bibinfo{author}{Hornreich, R.~M.} (\bibinfo{year}{1978}).
\newblock \bibinfo{title}{Magnetic interactions and weak ferromagnetism in the
  rare-earth orthochromites}.
\newblock {\it \bibinfo{journal}{J. Magn. Magn. Mater.}\/},  {\it
  \bibinfo{volume}{7}\/}, \bibinfo{pages}{280--285}.
\bibitem[{Huang et~al.(2021)Huang, Yi, Jiang, Kao, Yin, Beck, Korolkov,
  Proksch, Shieh \& Chen}]{HUANG2021116536}
\bibinfo{author}{Huang, K.-W.}, \bibinfo{author}{Yi, S.-H.},
  \bibinfo{author}{Jiang, Y.-S.}, \bibinfo{author}{Kao, W.-C.},
  \bibinfo{author}{Yin, Y.-T.}, \bibinfo{author}{Beck, D.},
  \bibinfo{author}{Korolkov, V.}, \bibinfo{author}{Proksch, R.},
  \bibinfo{author}{Shieh, J.}, \& \bibinfo{author}{Chen, M.-J.}
  (\bibinfo{year}{2021}).
\newblock \bibinfo{title}{Sub-7-nm textured {ZrO}$_2$ with giant
  ferroelectricity}.
\newblock {\it \bibinfo{journal}{Acta Materialia}\/},  {\it
  \bibinfo{volume}{205}\/}, \bibinfo{pages}{116536}.
\bibitem[{Hur et~al.(2004)Hur, Park, Sharma, Ahn, Guha \& Cheong}]{Hur2004}
\bibinfo{author}{Hur, N.}, \bibinfo{author}{Park, S.}, \bibinfo{author}{Sharma,
  P.~A.}, \bibinfo{author}{Ahn, J.~S.}, \bibinfo{author}{Guha, S.}, \&
  \bibinfo{author}{Cheong, S.-W.} (\bibinfo{year}{2004}).
\newblock \bibinfo{title}{Electric polarization reversal and memory in a
  multiferroic material induced by magnetic fields}.
\newblock {\it \bibinfo{journal}{Nature}\/},  {\it \bibinfo{volume}{429}\/},
  \bibinfo{pages}{392--395}.
\bibitem[{Jones et~al.(2014)Jones, Gaw, Doig, Prabhakaran,
  H$\acute{\textrm{e}}$troy~Wheeler, Boothroyd \& Lloyd-Hughes}]{Jones2014}
\bibinfo{author}{Jones, P.~P.}, \bibinfo{author}{Gaw, S.~M.},
  \bibinfo{author}{Doig, K.~I.}, \bibinfo{author}{Prabhakaran, D.},
  \bibinfo{author}{H$\acute{\textrm{e}}$troy~Wheeler, E.~M.},
  \bibinfo{author}{Boothroyd, A.~T.}, \& \bibinfo{author}{Lloyd-Hughes}
  (\bibinfo{year}{2014}).
\newblock \bibinfo{title}{High-temperature electromagnons in the magnetically
  induced multiferroic cupric oxide driven by intersublattice exchange}.
\newblock {\it \bibinfo{journal}{Nat. Commun.}\/},  {\it
  \bibinfo{volume}{5}\/}, \bibinfo{pages}{3787}.
\bibitem[{Ko et~al.(2007)Ko, Kim, Kim \& Choi}]{Ko2007}
\bibinfo{author}{Ko, E.}, \bibinfo{author}{Kim, B.~J.}, \bibinfo{author}{Kim,
  C.}, \& \bibinfo{author}{Choi, H.~J.} (\bibinfo{year}{2007}).
\newblock \bibinfo{title}{Strong orbital-dependent $d$-band hybridization and
  fermi-surface reconstruction in metallic {Ca}$_{2-x}${Sr}$_{x}${RuO}$_{4}$}.
\newblock {\it \bibinfo{journal}{Phys. Rev. Lett.}\/},  {\it
  \bibinfo{volume}{98}\/}, \bibinfo{pages}{226401}.
\bibitem[{Kresse \& Furthm{\"u}ller(1996)}]{Kresse1996}
\bibinfo{author}{Kresse, G.}, \& \bibinfo{author}{Furthm{\"u}ller, J.}
  (\bibinfo{year}{1996}).
\newblock \bibinfo{title}{Efficiency of ab-initio total energy calculations for
  metals and semiconductors using a plane-wave basis set}.
\newblock {\it \bibinfo{journal}{Compt. Mater. Sci.}\/},  {\it
  \bibinfo{volume}{6}\/}, \bibinfo{pages}{15--50}.
\bibitem[{Kresse \& Joubert(1999)}]{Kresse1999}
\bibinfo{author}{Kresse, G.}, \& \bibinfo{author}{Joubert, D.}
  (\bibinfo{year}{1999}).
\newblock \bibinfo{title}{From ultrasoft pseudopotentials to the projector
  augmented-wave method}.
\newblock {\it \bibinfo{journal}{Phys. Rev. B}\/},  {\it
  \bibinfo{volume}{59}\/}, \bibinfo{pages}{1758--1775}.
\bibitem[{Krynetskii \& Matveev(1997)}]{Krynetskii1997}
\bibinfo{author}{Krynetskii, I.~B.}, \& \bibinfo{author}{Matveev, V.~M.}
  (\bibinfo{year}{1997}).
\newblock \bibinfo{title}{Metamagnetism and magnetostriction of the ising
  antiferromagnet {DyCrO}$_{3}$}.
\newblock {\it \bibinfo{journal}{Phys. Solid State}\/},  {\it
  \bibinfo{volume}{39}\/}, \bibinfo{pages}{584--585}.
\bibitem[{Kumar et~al.(2016)Kumar, Xiao, Nair, Voigt, Schmitz, Chatterji,
  Jalarvo \& Br{\"u}ckel}]{Kumar2016}
\bibinfo{author}{Kumar, C. M.~N.}, \bibinfo{author}{Xiao, Y.},
  \bibinfo{author}{Nair, H.~S.}, \bibinfo{author}{Voigt, J.},
  \bibinfo{author}{Schmitz, B.}, \bibinfo{author}{Chatterji, T.},
  \bibinfo{author}{Jalarvo, N.~H.}, \& \bibinfo{author}{Br{\"u}ckel, T.}
  (\bibinfo{year}{2016}).
\newblock \bibinfo{title}{Hyperfine and crystal field interactions in
  multiferroic {HoCrO}$_3$}.
\newblock {\it \bibinfo{journal}{J. Phys.: Condens. Matter}\/},  {\it
  \bibinfo{volume}{28}\/}, \bibinfo{pages}{476001}.
\bibitem[{Landron \& Lepetit(2008)}]{Landron2008}
\bibinfo{author}{Landron, S.}, \& \bibinfo{author}{Lepetit, M.-B.}
  (\bibinfo{year}{2008}).
\newblock \bibinfo{title}{Importance of $t_{2g}$-${e}_{g}$ hybridization in
  transition metal oxides}.
\newblock {\it \bibinfo{journal}{Phys. Rev. B}\/},  {\it
  \bibinfo{volume}{77}\/}, \bibinfo{pages}{125106}.
\bibitem[{Li et~al.(2006)Li, Su, Persson, Meuffels, Walter, Skowronek \&
  Br{\"u}ckel}]{Li2006}
\bibinfo{author}{Li, H.}, \bibinfo{author}{Su, Y.}, \bibinfo{author}{Persson,
  J.}, \bibinfo{author}{Meuffels, P.}, \bibinfo{author}{Walter, J.},
  \bibinfo{author}{Skowronek, R.}, \& \bibinfo{author}{Br{\"u}ckel, T.}
  (\bibinfo{year}{2006}).
\newblock \bibinfo{title}{Correlation between structural and magnetic
  properties of {La}$_{7/8}${Sr}$_{1/8}${Mn}$_{1-\gamma}${O}$_{3+\delta}$ with
  controlled nonstoichiometry}.
\newblock {\it \bibinfo{journal}{Journal of Physics: Condensed Matter}\/},
  {\it \bibinfo{volume}{19}\/}, \bibinfo{pages}{016003}.
\bibitem[{Li et~al.(2007)Li, Su, Persson, Meuffels, Walter, Skowronek \&
  Br{\"u}ckel}]{Li2007_2}
\bibinfo{author}{Li, H.}, \bibinfo{author}{Su, Y.}, \bibinfo{author}{Persson,
  J.}, \bibinfo{author}{Meuffels, P.}, \bibinfo{author}{Walter, J.},
  \bibinfo{author}{Skowronek, R.}, \& \bibinfo{author}{Br{\"u}ckel, T.}
  (\bibinfo{year}{2007}).
\newblock \bibinfo{title}{Neutron-diffraction study of structural transition
  and magnetic order in orthorhombic and rhombohedral
  {La}$_{7/8}${Sr}$_{1/8}${Mn}$_{1-\gamma}${O}$_{3+\delta}$}.
\newblock {\it \bibinfo{journal}{Journal of Physics: Condensed Matter}\/},
  {\it \bibinfo{volume}{19}\/}, \bibinfo{pages}{176226}.
\bibitem[{Li et~al.(2021)Li, Zhu, Wu \& Tang}]{Zhu2019-1}
\bibinfo{author}{Li, H.}, \bibinfo{author}{Zhu, Y.}, \bibinfo{author}{Wu, S.},
  \& \bibinfo{author}{Tang, Z.} (\bibinfo{year}{2021}).
\newblock {\it \bibinfo{title}{A method of centimeter-sized single crystal
  growth of chromate compounds and related storage device}\/}.
\newblock \bibinfo{publisher}{China Patent CN110904497B}.
\bibitem[{Li(2008)}]{Li2008-1}
\bibinfo{author}{Li, H.-F.} (\bibinfo{year}{2008}).
\newblock \bibinfo{title}{Synthesis of {CMR} manganites and ordering phenomena
  in complex transition metal oxides}.
\newblock {\it \bibinfo{journal}{Forschungszentrum J\"{u}lich GmbH Press}\/},
  {\it \bibinfo{volume}{Ph.D. thesis}\/}.
\bibitem[{Li(2016)}]{HFLi2016}
\bibinfo{author}{Li, H.-F.} (\bibinfo{year}{2016}).
\newblock \bibinfo{title}{Possible ground states and parallel
  magnetic-field-driven phase transitions of collinear antiferromagnets}.
\newblock {\it \bibinfo{journal}{Npj Comput. Mater.}\/},  {\it
  \bibinfo{volume}{2}\/}, \bibinfo{pages}{1--8}.
\bibitem[{Li et~al.(2009)Li, Su, Xiao, Persson, Meuffels \&
  Br{\"u}ckel}]{Li2009}
\bibinfo{author}{Li, H.-F.}, \bibinfo{author}{Su, Y.}, \bibinfo{author}{Xiao,
  Y.}, \bibinfo{author}{Persson, J.}, \bibinfo{author}{Meuffels, P.}, \&
  \bibinfo{author}{Br{\"u}ckel, T.} (\bibinfo{year}{2009}).
\newblock \bibinfo{title}{Crystal and magnetic structure of single-crystal
  {La}$_{1-x}${Sr}$_x${Mn}{O}$_3$ ($x \approx$ 1/8)}.
\newblock {\it \bibinfo{journal}{The European Physical Journal B}\/},  {\it
  \bibinfo{volume}{67}\/}, \bibinfo{pages}{149--157}.
\bibitem[{Li et~al.(2018)Li, Gou, Li, Tian, Cong, Ju, Tian, Geng, Tan, Yang
  et~al.}]{li2018millimeter}
\bibinfo{author}{Li, Y.-T.}, \bibinfo{author}{Gou, G.-Y.}, \bibinfo{author}{Li,
  L.-S.}, \bibinfo{author}{Tian, H.}, \bibinfo{author}{Cong, X.},
  \bibinfo{author}{Ju, Z.-Y.}, \bibinfo{author}{Tian, Y.},
  \bibinfo{author}{Geng, X.-S.}, \bibinfo{author}{Tan, P.-H.},
  \bibinfo{author}{Yang, Y.} et~al. (\bibinfo{year}{2018}).
\newblock \bibinfo{title}{Millimeter-scale nonlocal photo-sensing based on
  single-crystal perovskite photodetector}.
\newblock {\it \bibinfo{journal}{IScience}\/},  {\it \bibinfo{volume}{7}\/},
  \bibinfo{pages}{110--119}.
\bibitem[{Monkhorst \& Pack(1976)}]{Monkhorst1976}
\bibinfo{author}{Monkhorst, H.~J.}, \& \bibinfo{author}{Pack, J.~D.}
  (\bibinfo{year}{1976}).
\newblock \bibinfo{title}{Special points for brillouin-zone integrations}.
\newblock {\it \bibinfo{journal}{Phys. Rev. B}\/},  {\it
  \bibinfo{volume}{13}\/}, \bibinfo{pages}{5188--5192}.
\bibitem[{Oliveira(2017)}]{Oliveira2017}
\bibinfo{author}{Oliveira, G.} (\bibinfo{year}{2017}).
\newblock \bibinfo{title}{Local probing spinel and perovskite complex magnetic
  systems}.
\newblock {\it \bibinfo{journal}{Universidade do Porto}\/},  {\it
  \bibinfo{volume}{Ph.D. thesis}\/}.
\bibitem[{Oliveira et~al.(2020)Oliveira, Teixeira, Moreira, Correia, Ara\'{u}jo
  \& Lopes}]{Oliveira2020}
\bibinfo{author}{Oliveira, G. N.~P.}, \bibinfo{author}{Teixeira, R.~C.},
  \bibinfo{author}{Moreira, R.~P.}, \bibinfo{author}{Correia, J.~G.},
  \bibinfo{author}{Ara\'{u}jo, J.~P.}, \& \bibinfo{author}{Lopes, A. M.~L.}
  (\bibinfo{year}{2020}).
\newblock \bibinfo{title}{Local inhomogeneous state in multiferroic
  {SmCrO}$_3$}.
\newblock {\it \bibinfo{journal}{Sci. Rep.}\/},  {\it \bibinfo{volume}{10}\/},
  \bibinfo{pages}{4686}.
\bibitem[{Ouladdiaf et~al.(2006)Ouladdiaf, Archer, McIntyre, Hewat, Brau \&
  York}]{Ouladdiaf2006}
\bibinfo{author}{Ouladdiaf, B.}, \bibinfo{author}{Archer, J.},
  \bibinfo{author}{McIntyre, G.~J.}, \bibinfo{author}{Hewat, A.~W.},
  \bibinfo{author}{Brau, D.}, \& \bibinfo{author}{York, S.}
  (\bibinfo{year}{2006}).
\newblock \bibinfo{title}{Orientexpress: A new system for laue neutron
  diffraction}.
\newblock {\it \bibinfo{journal}{Physica B}\/},  {\it \bibinfo{volume}{385}\/},
  \bibinfo{pages}{1052--1054}.
\bibitem[{Perdew et~al.(1996)Perdew, Burke \& Ernzerhof}]{Perdew1996}
\bibinfo{author}{Perdew, J.~P.}, \bibinfo{author}{Burke, K.}, \&
  \bibinfo{author}{Ernzerhof, M.} (\bibinfo{year}{1996}).
\newblock \bibinfo{title}{Generalized gradient approximation made simple}.
\newblock {\it \bibinfo{journal}{Phys. Rev. Lett.}\/},  {\it
  \bibinfo{volume}{77}\/}, \bibinfo{pages}{3865--3868}.
\bibitem[{Philipp et~al.(2003)Philipp, Majewski, Alff, Erb, Gross, Graf,
  Brandt, Simon, Walther, Mader, Topwal \& Sarma}]{Philipp2003}
\bibinfo{author}{Philipp, J.~B.}, \bibinfo{author}{Majewski, P.},
  \bibinfo{author}{Alff, L.}, \bibinfo{author}{Erb, A.},
  \bibinfo{author}{Gross, R.}, \bibinfo{author}{Graf, T.},
  \bibinfo{author}{Brandt, M.~S.}, \bibinfo{author}{Simon, J.},
  \bibinfo{author}{Walther, T.}, \bibinfo{author}{Mader, W.},
  \bibinfo{author}{Topwal, D.}, \& \bibinfo{author}{Sarma, D.~D.}
  (\bibinfo{year}{2003}).
\newblock \bibinfo{title}{Structural and doping effects in the half-metallic
  double perovskite {A}$_{2}${CrWO}$_{6}$ ({A = Sr, Ba, and Ca)}}.
\newblock {\it \bibinfo{journal}{Phys. Rev. B}\/},  {\it
  \bibinfo{volume}{68}\/}, \bibinfo{pages}{144431}.
\bibitem[{Preethi~Meher et~al.(2014)Preethi~Meher, Wahl, Maignan, Martin \&
  Lebedev}]{Meher2014}
\bibinfo{author}{Preethi~Meher, K. R.~S.}, \bibinfo{author}{Wahl, A.},
  \bibinfo{author}{Maignan, A.}, \bibinfo{author}{Martin, C.}, \&
  \bibinfo{author}{Lebedev, O.~I.} (\bibinfo{year}{2014}).
\newblock \bibinfo{title}{Observation of electric polarization reversal and
  magnetodielectric effect in orthochromites: A comparison between
  {LuCrO}$_{3}$ and {ErCrO}$_{3}$}.
\newblock {\it \bibinfo{journal}{Phys. Rev. B}\/},  {\it
  \bibinfo{volume}{89}\/}, \bibinfo{pages}{144401}.
\bibitem[{Qian et~al.(2014)Qian, Chen, Cao \& Zhang}]{Qian2014}
\bibinfo{author}{Qian, X.}, \bibinfo{author}{Chen, L.}, \bibinfo{author}{Cao,
  S.}, \& \bibinfo{author}{Zhang, J.} (\bibinfo{year}{2014}).
\newblock \bibinfo{title}{A study of the spin reorientation with t--e orbital
  hybridization in {SmCrO}$_{3}$}.
\newblock {\it \bibinfo{journal}{Solid State Commun.}\/},  {\it
  \bibinfo{volume}{195}\/}, \bibinfo{pages}{21--25}.
\bibitem[{Rajeswaran et~al.(2012)Rajeswaran, Khomskii, Zvezdin, Rao \&
  Sundaresan}]{Rajeswaran2012}
\bibinfo{author}{Rajeswaran, B.}, \bibinfo{author}{Khomskii, D.~I.},
  \bibinfo{author}{Zvezdin, A.~K.}, \bibinfo{author}{Rao, C. N.~R.}, \&
  \bibinfo{author}{Sundaresan, A.} (\bibinfo{year}{2012}).
\newblock \bibinfo{title}{Field-induced polar order at the {N\'eel} temperature
  of chromium in rare-earth orthochromites: Interplay of rare-earth and {Cr}
  magnetism}.
\newblock {\it \bibinfo{journal}{Phys. Rev. B}\/},  {\it
  \bibinfo{volume}{86}\/}, \bibinfo{pages}{214409}.
\bibitem[{Serrao et~al.(2005)Serrao, Kundu, Krupanidhi, Waghmare \&
  Rao}]{Serrao2005}
\bibinfo{author}{Serrao, C.~R.}, \bibinfo{author}{Kundu, A.~K.},
  \bibinfo{author}{Krupanidhi, S.~B.}, \bibinfo{author}{Waghmare, U.~V.}, \&
  \bibinfo{author}{Rao, C. N.~R.} (\bibinfo{year}{2005}).
\newblock \bibinfo{title}{Biferroic {YCrO}$_{3}$}.
\newblock {\it \bibinfo{journal}{Phys. Rev. B}\/},  {\it
  \bibinfo{volume}{72}\/}, \bibinfo{pages}{220101}.
\bibitem[{Shamir et~al.(1981)Shamir, Shaked \& Shtrikman}]{Shamir1981}
\bibinfo{author}{Shamir, N.}, \bibinfo{author}{Shaked, H.}, \&
  \bibinfo{author}{Shtrikman, S.} (\bibinfo{year}{1981}).
\newblock \bibinfo{title}{Magnetic structure of some rare-earth
  orthochromites}.
\newblock {\it \bibinfo{journal}{Phys. Rev. B}\/},  {\it
  \bibinfo{volume}{24}\/}, \bibinfo{pages}{6642--6651}.
\bibitem[{Shi et~al.(2018)Shi, Yin, Seehra \& Jain}]{Shi2018}
\bibinfo{author}{Shi, J.}, \bibinfo{author}{Yin, S.}, \bibinfo{author}{Seehra,
  M.~S.}, \& \bibinfo{author}{Jain, M.} (\bibinfo{year}{2018}).
\newblock \bibinfo{title}{Enhancement in magnetocaloric properties of
  {ErCrO}$_{3}$ via {A}-{site} {Gd} substitution}.
\newblock {\it \bibinfo{journal}{J. Appl. Phys.}\/},  {\it
  \bibinfo{volume}{123}\/}, \bibinfo{pages}{193901}.
\bibitem[{Shick et~al.(1999)Shick, Liechtenstein \& Pickett}]{Shick1999}
\bibinfo{author}{Shick, A.~B.}, \bibinfo{author}{Liechtenstein, A.~I.}, \&
  \bibinfo{author}{Pickett, W.~E.} (\bibinfo{year}{1999}).
\newblock \bibinfo{title}{Implementation of the {LDA}{+}{U} method using the
  full-potential linearized augmented plane-wave basis}.
\newblock {\it \bibinfo{journal}{Phys. Rev. B}\/},  {\it
  \bibinfo{volume}{60}\/}, \bibinfo{pages}{10763--10769}.
\bibitem[{Siddique et~al.(2021)Siddique, Faizan, Riyajuddin, Tripathi, Ahmad \&
  Ghosh}]{Siddiqu2021}
\bibinfo{author}{Siddique, M.~N.}, \bibinfo{author}{Faizan, M.},
  \bibinfo{author}{Riyajuddin, S.}, \bibinfo{author}{Tripathi, P.},
  \bibinfo{author}{Ahmad, S.}, \& \bibinfo{author}{Ghosh, K.}
  (\bibinfo{year}{2021}).
\newblock \bibinfo{title}{Intrinsic structural distortion assisted optical and
  magnetic properties of orthorhombic rare-earth perovskite
  {La}$_{1-x}${Eu}$_{x}${CrO}$_{3}$: Effect of te hybridization}.
\newblock {\it \bibinfo{journal}{J. Alloys Compd.}\/},  {\it
  \bibinfo{volume}{850}\/}, \bibinfo{pages}{156748}.
\bibitem[{Singh et~al.(2018)Singh, Pandit \& Kumar}]{Singh2018}
\bibinfo{author}{Singh, K.~D.}, \bibinfo{author}{Pandit, R.}, \&
  \bibinfo{author}{Kumar, R.} (\bibinfo{year}{2018}).
\newblock \bibinfo{title}{Effect of rare earth ions on structural and optical
  properties of specific perovskite orthochromates; {RCrO}$_{3}$ ({R = La, Nd,
  Eu, Gd, Dy, and Y})}.
\newblock {\it \bibinfo{journal}{Solid State Sci.}\/},  {\it
  \bibinfo{volume}{85}\/}, \bibinfo{pages}{70--75}.
\bibitem[{Slater \& Koster(1954)}]{Slater1954}
\bibinfo{author}{Slater, J.~C.}, \& \bibinfo{author}{Koster, G.~F.}
  (\bibinfo{year}{1954}).
\newblock \bibinfo{title}{Simplified {LCAO} method for the periodic potential
  problem}.
\newblock {\it \bibinfo{journal}{Phys. Rev.}\/},  {\it \bibinfo{volume}{94}\/},
  \bibinfo{pages}{1498--1524}.
\bibitem[{Spaldin \& Ramesh(2019)}]{Spaldin2019}
\bibinfo{author}{Spaldin, N.~A.}, \& \bibinfo{author}{Ramesh, R.}
  (\bibinfo{year}{2019}).
\newblock \bibinfo{title}{Advances in magnetoelectric multiferroics}.
\newblock {\it \bibinfo{journal}{Nat. Mater.}\/},  {\it
  \bibinfo{volume}{18}\/}, \bibinfo{pages}{203--212}.
\bibitem[{Streltsov \& Khomskii(2008)}]{Streltsov2008}
\bibinfo{author}{Streltsov, S.~V.}, \& \bibinfo{author}{Khomskii, D.~I.}
  (\bibinfo{year}{2008}).
\newblock \bibinfo{title}{Electronic structure and magnetic properties of
  pyroxenes {(Li,Na)TM(Si,Ge)}$_2${O}$_6$: {Low}-dimensional magnets with
  90\ifmmode^\circ\else\textdegree\fi{} bonds}.
\newblock {\it \bibinfo{journal}{Phys. Rev. B}\/},  {\it
  \bibinfo{volume}{77}\/}, \bibinfo{pages}{064405}.
\bibitem[{Su et~al.(2010{\natexlab{a}})Su, Zhang, Feng, Li, Li, Zhou, Chen \&
  Cao}]{Su2010-2}
\bibinfo{author}{Su, Y.}, \bibinfo{author}{Zhang, J.}, \bibinfo{author}{Feng,
  Z.}, \bibinfo{author}{Li, L.}, \bibinfo{author}{Li, B.},
  \bibinfo{author}{Zhou, Y.}, \bibinfo{author}{Chen, Z.}, \&
  \bibinfo{author}{Cao, S.} (\bibinfo{year}{2010}{\natexlab{a}}).
\newblock \bibinfo{title}{Magnetization reversal and {Yb}$^{3+}$/{Cr}$^{3+}$
  spin ordering at low temperature for perovskite {YbCrO}$_{3}$ chromites}.
\newblock {\it \bibinfo{journal}{J. Appl. Phys.}\/},  {\it
  \bibinfo{volume}{108}\/}, \bibinfo{pages}{013905}.
\bibitem[{Su et~al.(2011)Su, Zhang, Feng, Li, Yan \& Cao}]{Su2011}
\bibinfo{author}{Su, Y.}, \bibinfo{author}{Zhang, J.}, \bibinfo{author}{Feng,
  Z.}, \bibinfo{author}{Li, Z.}, \bibinfo{author}{Yan, S.}, \&
  \bibinfo{author}{Cao, S.} (\bibinfo{year}{2011}).
\newblock \bibinfo{title}{Magnetic properties of rare earth {HoCrO}$_{3}$
  chromites}.
\newblock {\it \bibinfo{journal}{J. Rare Earth.}\/},  {\it
  \bibinfo{volume}{29}\/}, \bibinfo{pages}{1060--1065}.
\bibitem[{Su et~al.(2012)Su, Zhang, Li, Kang, Yu, Jing \& Cao}]{Su2012}
\bibinfo{author}{Su, Y.}, \bibinfo{author}{Zhang, J.}, \bibinfo{author}{Li,
  B.}, \bibinfo{author}{Kang, B.}, \bibinfo{author}{Yu, Q.},
  \bibinfo{author}{Jing, C.}, \& \bibinfo{author}{Cao, S.}
  (\bibinfo{year}{2012}).
\newblock \bibinfo{title}{The dependence of magnetic properties on temperature
  for rare earth {ErCrO}$_{3}$ chromites}.
\newblock {\it \bibinfo{journal}{Ceram. Int.}\/},  {\it
  \bibinfo{volume}{38}\/}, \bibinfo{pages}{S421--S424}.
\bibitem[{Su et~al.(2010{\natexlab{b}})Su, Zhang, Li, Li, Zhou, Deng, Chen \&
  Cao}]{Su2010}
\bibinfo{author}{Su, Y.}, \bibinfo{author}{Zhang, J.}, \bibinfo{author}{Li,
  L.}, \bibinfo{author}{Li, B.}, \bibinfo{author}{Zhou, Y.},
  \bibinfo{author}{Deng, D.}, \bibinfo{author}{Chen, Z.}, \&
  \bibinfo{author}{Cao, S.} (\bibinfo{year}{2010}{\natexlab{b}}).
\newblock \bibinfo{title}{Temperature dependence of magnetic properties and
  change of specific heat in perovskite {ErCrO}$_{3}$ chromites}.
\newblock {\it \bibinfo{journal}{Appl. Phys. A}\/},  {\it
  \bibinfo{volume}{100}\/}, \bibinfo{pages}{73--78}.
\bibitem[{Sun et~al.(2019)Sun, Deng, Xu, Xu, Li, Wu, Qian, Zhong, Nuckolls,
  Harutyunyan et~al.}]{sun2019anisotropic}
\bibinfo{author}{Sun, D.}, \bibinfo{author}{Deng, G.-H.}, \bibinfo{author}{Xu,
  B.}, \bibinfo{author}{Xu, E.}, \bibinfo{author}{Li, X.}, \bibinfo{author}{Wu,
  Y.}, \bibinfo{author}{Qian, Y.}, \bibinfo{author}{Zhong, Y.},
  \bibinfo{author}{Nuckolls, C.}, \bibinfo{author}{Harutyunyan, A.~R.} et~al.
  (\bibinfo{year}{2019}).
\newblock \bibinfo{title}{Anisotropic singlet fission in single crystalline
  hexacene}.
\newblock {\it \bibinfo{journal}{Iscience}\/},  {\it \bibinfo{volume}{19}\/},
  \bibinfo{pages}{1079--1089}.
\bibitem[{Taheri et~al.(2016)Taheri, Razavi, Yamani, Flacau, Reuvekamp, Schulz
  \& Kremer}]{Taheri2016}
\bibinfo{author}{Taheri, M.}, \bibinfo{author}{Razavi, F.~S.},
  \bibinfo{author}{Yamani, Z.}, \bibinfo{author}{Flacau, R.},
  \bibinfo{author}{Reuvekamp, P.~G.}, \bibinfo{author}{Schulz, A.}, \&
  \bibinfo{author}{Kremer, R.~K.} (\bibinfo{year}{2016}).
\newblock \bibinfo{title}{Magnetic structure, magnetoelastic coupling, and
  thermal properties of {EuCrO}$_{3}$ nanopowders}.
\newblock {\it \bibinfo{journal}{Phys. Rev. B}\/},  {\it
  \bibinfo{volume}{93}\/}, \bibinfo{pages}{104414}.
\bibitem[{Tamaki et~al.(1977)Tamaki, Tsushima \& Yamaguchi}]{Tamaki1977}
\bibinfo{author}{Tamaki, T.}, \bibinfo{author}{Tsushima, K.}, \&
  \bibinfo{author}{Yamaguchi, Y.} (\bibinfo{year}{1977}).
\newblock \bibinfo{title}{Spin reorientation in {TmCrO}$_{3}$}.
\newblock {\it \bibinfo{journal}{Physica B+C}\/},  {\it
  \bibinfo{volume}{86}\/}, \bibinfo{pages}{923--924}.
\bibitem[{Tsushima et~al.(1974)Tsushima, Tamaki \& Yamaura}]{Tsushima1974}
\bibinfo{author}{Tsushima, K.}, \bibinfo{author}{Tamaki, T.}, \&
  \bibinfo{author}{Yamaura, R.} (\bibinfo{year}{1974}).
\newblock \bibinfo{title}{Proceedings of the international conference on
  magnetism, {Moscow}, 1973}.
\newblock (p. \bibinfo{pages}{270}).
\newblock \bibinfo{publisher}{Nauka, Moscow} volume~\bibinfo{volume}{5}.
\bibitem[{Vagadia et~al.(2018)Vagadia, Rayaprol \& Nigam}]{Vagadia2018}
\bibinfo{author}{Vagadia, M.}, \bibinfo{author}{Rayaprol, S.}, \&
  \bibinfo{author}{Nigam, A.} (\bibinfo{year}{2018}).
\newblock \bibinfo{title}{Influence of mn-substitution on the magnetic and
  thermal properties of {TbCrO}$_3$}.
\newblock {\it \bibinfo{journal}{J. Alloys Compd.}\/},  {\it
  \bibinfo{volume}{735}\/}, \bibinfo{pages}{1031--1040}.
\bibitem[{Wang et~al.(2016{\natexlab{a}})Wang, Rao, Zhang, Zhang, Wang \&
  Yao}]{Wang2016-1}
\bibinfo{author}{Wang, L.}, \bibinfo{author}{Rao, G.~H.},
  \bibinfo{author}{Zhang, X.}, \bibinfo{author}{Zhang, L.~L.},
  \bibinfo{author}{Wang, S.~W.}, \& \bibinfo{author}{Yao, Q.~R.}
  (\bibinfo{year}{2016}{\natexlab{a}}).
\newblock \bibinfo{title}{Reversals of magnetization and exchange-bias in
  perovskite chromite {TmCrO}$_{3}$}.
\newblock {\it \bibinfo{journal}{Ceram. Int.}\/},  {\it
  \bibinfo{volume}{42}\/}, \bibinfo{pages}{10171--10174}.
\bibitem[{Wang et~al.(2016{\natexlab{b}})Wang, Wang, Zhang, Zhang, Yao \&
  Rao}]{Wang2016-2}
\bibinfo{author}{Wang, L.}, \bibinfo{author}{Wang, S.~W.},
  \bibinfo{author}{Zhang, X.}, \bibinfo{author}{Zhang, L.~L.},
  \bibinfo{author}{Yao, R.}, \& \bibinfo{author}{Rao, G.~H.}
  (\bibinfo{year}{2016}{\natexlab{b}}).
\newblock \bibinfo{title}{Reversals of magnetization and exchange-bias in
  perovskite chromite {YbCrO}$_{3}$}.
\newblock {\it \bibinfo{journal}{J. Alloys Compd.}\/},  {\it
  \bibinfo{volume}{662}\/}, \bibinfo{pages}{268--271}.
\bibitem[{Wu et~al.(2020)Wu, Zhu, Gao, Xiao, Xia, Zhou, Ouyang, Li, Chen, Tang
  et~al.}]{Wu2020}
\bibinfo{author}{Wu, S.}, \bibinfo{author}{Zhu, Y.}, \bibinfo{author}{Gao, H.},
  \bibinfo{author}{Xiao, Y.}, \bibinfo{author}{Xia, J.}, \bibinfo{author}{Zhou,
  P.}, \bibinfo{author}{Ouyang, D.}, \bibinfo{author}{Li, Z.},
  \bibinfo{author}{Chen, Z.}, \bibinfo{author}{Tang, Z.} et~al.
  (\bibinfo{year}{2020}).
\newblock \bibinfo{title}{Super-necking crystal growth and structural and
  magnetic properties of {SrTb}$_{2}${O}$_{4}$ single crystals}.
\newblock {\it \bibinfo{journal}{ACS omega}\/},  {\it \bibinfo{volume}{5}\/},
  \bibinfo{pages}{16584--16594}.
\bibitem[{Xiong et~al.(2021)Xiong, Peng \& Yang}]{xiong2021near}
\bibinfo{author}{Xiong, P.}, \bibinfo{author}{Peng, M.}, \&
  \bibinfo{author}{Yang, Z.} (\bibinfo{year}{2021}).
\newblock \bibinfo{title}{Near-infrared mechanoluminescence crystals: a
  review}.
\newblock {\it \bibinfo{journal}{Iscience}\/},  {\it \bibinfo{volume}{24}\/}.
\bibitem[{Yaresko(2008)}]{Yaresko2008}
\bibinfo{author}{Yaresko, A.~N.} (\bibinfo{year}{2008}).
\newblock \bibinfo{title}{Electronic band structure and exchange coupling
  constants in ${A}${Cr}$_2{X}_{4}$ spinels (${A}$ = {Zn, Cd, Hg}; ${X}$ = {O,
  S, Se})}.
\newblock {\it \bibinfo{journal}{Phys. Rev. B}\/},  {\it
  \bibinfo{volume}{77}\/}, \bibinfo{pages}{115106}.
\bibitem[{Yekta et~al.(2021)Yekta, Hadipour, \ifmmode \mbox{\c{S}}\else
  \c{S}\fi{}a\ifmmode \mbox{\c{s}}\else \c{s}\fi{}\ifmmode \imath \else \i
  \fi{}o\ifmmode~\breve{g}\else \u{g}\fi{}lu, Friedrich, Jafari, Bl\"ugel \&
  Mertig}]{Yekta2021}
\bibinfo{author}{Yekta, Y.}, \bibinfo{author}{Hadipour, H.},
  \bibinfo{author}{\ifmmode \mbox{\c{S}}\else \c{S}\fi{}a\ifmmode
  \mbox{\c{s}}\else \c{s}\fi{}\ifmmode \imath \else \i
  \fi{}o\ifmmode~\breve{g}\else \u{g}\fi{}lu, E.}, \bibinfo{author}{Friedrich,
  C.}, \bibinfo{author}{Jafari, S.~A.}, \bibinfo{author}{Bl\"ugel, S.}, \&
  \bibinfo{author}{Mertig, I.} (\bibinfo{year}{2021}).
\newblock \bibinfo{title}{Strength of effective coulomb interaction in
  two-dimensional transition-metal halides ${MX}_{2}$ and ${MX}_{3}$ (${M}$ =
  {Ti, V, Cr, Mn, Fe, Co, Ni}; ${X}$ = {Cl, Br, I})}.
\newblock {\it \bibinfo{journal}{Phys. Rev. Mater.}\/},  {\it
  \bibinfo{volume}{5}\/}, \bibinfo{pages}{034001}.
\bibitem[{Yin et~al.(2016)Yin, Yang, Tong, Luo, Park, Shin, Song, Dai, Kim, Zhu
  et~al.}]{Yin2016}
\bibinfo{author}{Yin, L.}, \bibinfo{author}{Yang, J.}, \bibinfo{author}{Tong,
  P.}, \bibinfo{author}{Luo, X.}, \bibinfo{author}{Park, C.},
  \bibinfo{author}{Shin, K.}, \bibinfo{author}{Song, W.}, \bibinfo{author}{Dai,
  J.}, \bibinfo{author}{Kim, K.}, \bibinfo{author}{Zhu, X.} et~al.
  (\bibinfo{year}{2016}).
\newblock \bibinfo{title}{Role of rare earth ions in the magnetic,
  magnetocaloric and magnetoelectric properties of {RCrO}$_3$ ({R} {=} {Dy},
  {Nd}, {Tb}, {Er}) crystals}.
\newblock {\it \bibinfo{journal}{J. Mater. Chem. C}\/},  {\it
  \bibinfo{volume}{4}\/}, \bibinfo{pages}{11198--11204}.
\bibitem[{Yin et~al.(2018)Yin, Shi, Zhang, Park, Kim, Yang, Tong, Song, Dai,
  Zhu, Yan \& Sun}]{Yin2018}
\bibinfo{author}{Yin, L.~H.}, \bibinfo{author}{Shi, T.~F.},
  \bibinfo{author}{Zhang, R.~R.}, \bibinfo{author}{Park, C.~B.},
  \bibinfo{author}{Kim, K.~H.}, \bibinfo{author}{Yang, J.},
  \bibinfo{author}{Tong, P.}, \bibinfo{author}{Song, W.~H.},
  \bibinfo{author}{Dai, J.~M.}, \bibinfo{author}{Zhu, X.~B.},
  \bibinfo{author}{Yan, W.~S.}, \& \bibinfo{author}{Sun, Y.~P.}
  (\bibinfo{year}{2018}).
\newblock \bibinfo{title}{Electric dipoles via {Cr}$^{3+}({d}^{3})$ ion
  off-center displacement in perovskite {DyCrO}$_{3}$}.
\newblock {\it \bibinfo{journal}{Phys. Rev. B}\/},  {\it
  \bibinfo{volume}{98}\/}, \bibinfo{pages}{054301}.
\bibitem[{Yin et~al.(2015)Yin, Yang, Kan, Song, Dai \& Sun}]{Yin2015}
\bibinfo{author}{Yin, L.~H.}, \bibinfo{author}{Yang, J.}, \bibinfo{author}{Kan,
  X.~C.}, \bibinfo{author}{Song, W.~H.}, \bibinfo{author}{Dai, J.~M.}, \&
  \bibinfo{author}{Sun, Y.~P.} (\bibinfo{year}{2015}).
\newblock \bibinfo{title}{Giant magnetocaloric effect and temperature induced
  magnetization jump in {GdCrO}$_3$ single crystal}.
\newblock {\it \bibinfo{journal}{J. Appl. Phys.}\/},  {\it
  \bibinfo{volume}{117}\/}, \bibinfo{pages}{133901}.
\bibitem[{Yoshii(2012)}]{Yoshii2012}
\bibinfo{author}{Yoshii, K.} (\bibinfo{year}{2012}).
\newblock \bibinfo{title}{Magnetization reversal in {TmCrO}$_{3}$}.
\newblock {\it \bibinfo{journal}{Mater. Res. Bull.}\/},  {\it
  \bibinfo{volume}{47}\/}, \bibinfo{pages}{3243--3248}.
\bibitem[{Yoshii \& Ikeda(2019)}]{Yoshii2019}
\bibinfo{author}{Yoshii, K.}, \& \bibinfo{author}{Ikeda, N.}
  (\bibinfo{year}{2019}).
\newblock \bibinfo{title}{Dielectric and magnetocaloric study of {TmCrO}$_3$}.
\newblock {\it \bibinfo{journal}{J. Alloys Compd.}\/},  {\it
  \bibinfo{volume}{804}\/}, \bibinfo{pages}{364--369}.
\bibitem[{Zhang(2020)}]{ZHANG20201694}
\bibinfo{author}{Zhang, Y.} (\bibinfo{year}{2020}).
\newblock \bibinfo{title}{Building a library of metre-scale high-index
  single-crystal copper foils}.
\newblock {\it \bibinfo{journal}{Science Bulletin}\/},  {\it
  \bibinfo{volume}{65}\/}, \bibinfo{pages}{1694--1695}.
\bibitem[{Zhou et~al.(2010)Zhou, Alonso, Pomjakushin, Goodenough, Ren, Yan \&
  Cheng}]{Zhou2010}
\bibinfo{author}{Zhou, J.-S.}, \bibinfo{author}{Alonso, J.~A.},
  \bibinfo{author}{Pomjakushin, V.}, \bibinfo{author}{Goodenough, J.~B.},
  \bibinfo{author}{Ren, Y.}, \bibinfo{author}{Yan, J.-Q.}, \&
  \bibinfo{author}{Cheng, J.-G.} (\bibinfo{year}{2010}).
\newblock \bibinfo{title}{Intrinsic structural distortion and superexchange
  interaction in the orthorhombic rare-earth perovskites {RCrO}$_{3}$}.
\newblock {\it \bibinfo{journal}{Phys. Rev. B}\/},  {\it
  \bibinfo{volume}{81}\/}, \bibinfo{pages}{214115}.
\bibitem[{Zhou \& Goodenough(2008)}]{Zhou2008}
\bibinfo{author}{Zhou, J.-S.}, \& \bibinfo{author}{Goodenough, J.~B.}
  (\bibinfo{year}{2008}).
\newblock \bibinfo{title}{Intrinsic structural distortion in orthorhombic
  perovskite oxides}.
\newblock {\it \bibinfo{journal}{Phys. Rev. B}\/},  {\it
  \bibinfo{volume}{77}\/}, \bibinfo{pages}{132104}.
\bibitem[{Zhu et~al.(2020{\natexlab{a}})Zhu, Fu, Tu, Li, Miao, Zhao, Wu, Xia,
  Zhou, Huq, Schmidt, Ouyang, Tang, He \& Li}]{Zhu2020-2}
\bibinfo{author}{Zhu, Y.}, \bibinfo{author}{Fu, Y.}, \bibinfo{author}{Tu, B.},
  \bibinfo{author}{Li, T.}, \bibinfo{author}{Miao, J.}, \bibinfo{author}{Zhao,
  Q.}, \bibinfo{author}{Wu, S.}, \bibinfo{author}{Xia, J.},
  \bibinfo{author}{Zhou, P.}, \bibinfo{author}{Huq, A.},
  \bibinfo{author}{Schmidt, W.}, \bibinfo{author}{Ouyang, D.},
  \bibinfo{author}{Tang, Z.}, \bibinfo{author}{He, Z.}, \& \bibinfo{author}{Li,
  H.-F.} (\bibinfo{year}{2020}{\natexlab{a}}).
\newblock \bibinfo{title}{Crystalline and magnetic structures, magnetization,
  heat capacity, and anisotropic magnetostriction effect in a yttrium-chromium
  oxide}.
\newblock {\it \bibinfo{journal}{Phys. Rev. Mater.}\/},  {\it
  \bibinfo{volume}{4}\/}, \bibinfo{pages}{094409}.
\bibitem[{Zhu et~al.(2020{\natexlab{b}})Zhu, Wu, Tu, Jin, Huq, Persson, Gao,
  Ouyang, He, Yao, Tang \& Li}]{Zhu2020-1}
\bibinfo{author}{Zhu, Y.}, \bibinfo{author}{Wu, S.}, \bibinfo{author}{Tu, B.},
  \bibinfo{author}{Jin, S.}, \bibinfo{author}{Huq, A.},
  \bibinfo{author}{Persson, J.}, \bibinfo{author}{Gao, H.},
  \bibinfo{author}{Ouyang, D.}, \bibinfo{author}{He, Z.}, \bibinfo{author}{Yao,
  D.-X.}, \bibinfo{author}{Tang, Z.}, \& \bibinfo{author}{Li, H.-F.}
  (\bibinfo{year}{2020}{\natexlab{b}}).
\newblock \bibinfo{title}{High-temperature magnetism and crystallography of a
  {YCrO}$_{3}$ single crystal}.
\newblock {\it \bibinfo{journal}{Phys. Rev. B}\/},  {\it
  \bibinfo{volume}{101}\/}, \bibinfo{pages}{014114}.
\bibitem[{Zhu et~al.(2020{\natexlab{c}})Zhu, Zhou, Li, Xia, Wu, Fu, Sun, Zhao,
  Li, Tang, Xiao, Chen \& Li}]{Zhu2020-3}
\bibinfo{author}{Zhu, Y.}, \bibinfo{author}{Zhou, P.}, \bibinfo{author}{Li,
  T.}, \bibinfo{author}{Xia, J.}, \bibinfo{author}{Wu, S.},
  \bibinfo{author}{Fu, Y.}, \bibinfo{author}{Sun, K.}, \bibinfo{author}{Zhao,
  Q.}, \bibinfo{author}{Li, Z.}, \bibinfo{author}{Tang, Z.},
  \bibinfo{author}{Xiao, Y.}, \bibinfo{author}{Chen, Z.}, \&
  \bibinfo{author}{Li, H.-F.} (\bibinfo{year}{2020}{\natexlab{c}}).
\newblock \bibinfo{title}{Enhanced magnetocaloric effect and magnetic phase
  diagrams of single-crystal {GdCrO}$_{3}$}.
\newblock {\it \bibinfo{journal}{Phys. Rev. B}\/},  {\it
  \bibinfo{volume}{102}\/}, \bibinfo{pages}{144425}.
\bibitem[{Van~der Ziel \& Van~Uitert(1969)}]{Ziel1969}
\bibinfo{author}{Van~der Ziel, J.~P.}, \& \bibinfo{author}{Van~Uitert, L.~G.}
  (\bibinfo{year}{1969}).
\newblock \bibinfo{title}{Magnon-assisted optical emission in {YCrO}$_{3}$ and
  {LuCrO}$_{3}$}.
\newblock {\it \bibinfo{journal}{Phys. Rev.}\/},  {\it
  \bibinfo{volume}{179}\/}, \bibinfo{pages}{343--351}.

\end{thebibliography}

\clearpage

\section*{STAR$\star$METHODS}

\section*{RESOURCES AVAILABILITY}

\section*{Lead contact}
Further information and requests for resources should be directed to and will be fulfilled by the lead contact, Prof. Dr. Hai-Feng Li (haifengli@um.edu.mo)

\section*{Materials availability}
This long-term project produced a series of RECrO$_3$ (RE = Y, Eu, Gd, Tb, Dy, Ho, Er, Tm, Yb, and Lu) single crystals utilizing the innovative method described in the China Invention Patent (CN110904497B). We welcome potential collaborations.

\section*{Data and code availability}
$\bullet$ All data reported in this article will be shared by the lead contact upon request. \newline
$\bullet$ Code with instructions reported in this article will be shared by the lead contact upon request. \newline
$\bullet$ Any additional information required to reanalyse the data reported in this study is available from the
lead contact upon request. \newline

\section*{METHOD DETAILS}
\subsection*{Single crystal growth procedure and parameters}

Using raw materials of Y$_2$O$_3$ (Alfa Aesar, 99.9\%), Gd$_2$O$_3$ (Alfa Aesar, 99.9\%), Tb$_4$O$_7$ (Alfa Aesar, 99.9\%), Dy$_2$O$_3$ (Alfa Aesar, 99.9\%), Ho$_2$O$_3$ (Alfa Aesar, 99.9\%), Er$_2$O$_3$ (Alfa Aesar, 99.9\%), Tm$_2$O$_3$ (Alfa Aesar, 99.9\%), Yb$_2$O$_3$ (Alfa Aesar, 99.9\%), Lu$_2$O$_3$ (Alfa Aesar, 99.9\%), and Cr$_2$O$_3$ (Alfa Aesar, 99.6\%), polycrystalline RECrO$_3$ samples were synthesized with solid-state reactions. We first obtained homogeneous polycrystalline powder with a single phase. After that, cylindrical feed rods with additional treatments \citep{Zhu2019-1} were shaped by a hydrostatic pressure of $\sim$ 70 MPa \citep{Wu2020, Zhu2020-1}. Then RECrO$_3$ single crystals were grown by a laser-diode floating-zone furnace (Model: LD-FZ-5-200W-VPO-PCUM). The floating-zone method assures no introduction of impurities \citep{Li2008-1}. The growth speed was fixed at 5--15 mm/h to attain a stable growth state. Due to the intense volatility of chromium oxides, we added extra 5--15\% mole raw chromium oxide for the synthesis of polycrystalline samples and for the growth of single crystals.

\subsection*{Neutron Laue diffraction}

To determine the quality of the grown single crystals, we performed a neutron Laue diffraction study on the diffractometer, OrientExpress, located at ILL, Grenoble, France. Simultaneously, we simulated the recorded neutron Laue patterns along the three crystallograhpic axes with the software of OrientExpress \citep{Ouladdiaf2006} to confirm the quality of the grown crystals.

\subsection*{Magnetization measurements}

Magnetization was measured using the option of a vibrating sample magnetometer of Quantum Design physical property measurement system. Small RECrO$_3$ single crystals (5--15 mg) were glued on a quartz sample holder with GE Varnish. The dc magnetization was measured at applied magnetic fields of 0, 50, and 100 Oe with zero-field cooling and field cooling modes in the temperature range of 1.8--400 K. The magnetic field dependent hysteresis loops were measured from $-$14 to 14 T at different temperatures within 1.8--300 K.

\subsection*{First-principles calculations}

The first-principles calculations of RECrO$_3$ compounds were carried out within density functional theory. The exchange and correlation term in Kohn-Sham equation was treated with the Perdew-Burke-Ernzerhof (PBE) and (PBE + $U$) functionals \citep{Perdew1996, Anisimov1997, Shick1999, Franchini2007} using the Vienna Ab-initio Simulation Package \citep{Kresse1996}. The core electrons were frozen, and the projected-augmented-wave method was used \citep{Kresse1999}. The Cr 3\emph{d}4\emph{s}, RE (RE = Eu--Lu) elements 5\emph{p}5\emph{d}6\emph{s}, Y $(4s4p5s4d)$, and O $2s2p$ electrons were treated as valence electrons. It is well known that the 4\emph{f} orbitals are tightly localized in comparison to the \emph{d} orbitals. In the present study, we aim mainly to unravel the effect of lattice variation on Cr$^{3+}$-O$^{2-}$-Cr$^{3+}$ superexchange interactions, and the energy scale of RE$^{3+}$-RE$^{3+}$ exchange interactions is about two magnitudes smaller than that of Cr$^{3+}$-Cr$^{3+}$, hence the 4\emph{f} electrons of lanthanide ions could be frozen reasonably. A Gaussian broadening of 0.05 eV was chosen. The set of plane-wave basis with an energy cutoff of 500 eV was used. Brillouin-zone integrations were performed with a Gamma-point-centered 7$\times$7$\times$5 Monkhorst-Pack k-point mesh \citep{Monkhorst1976}. The ionic relaxation was performed with a convergency criterion of 10$^{-5}$ eV/primitive cell for each relaxation step and stopped moving when residual force $<$ 0.01 eV/{\AA}. A convergency accuracy of 10$^{-6}$ eV per conventional cell was chosen for subsequent static self-consistent calculations.

We first optimized the structures with collinear magnetic configurations of FM and \emph{A}-type, \emph{C}-type, and \emph{G}-type AFM, to determine the magnetic ground state of each compound. Furthermore, to avoid an underestimation of the band gap, the Hubbard $U$ value ($U_\texttt{eff} = U - J$) of each compound was calculated using the linear response ansatz \citep{Dudarev1998, Cococcioni2005}. With appropriate Hubbard $U$ value, static electronic self-consistent calculations were performed using the (PBE + $U$) method to obtain the exact total energy of the four magnetic states. The exchange parameters, $J_i$, under classical Heisenberg model could be deduced with the energy mapping method, as does $T^{\texttt{Cr}}_\texttt{N}$ based on the mean-field approximation (MFA). Finally, the electronic structures of the RECrO$_3$ compounds were calculated with reading the charge density of ground state of the magnetic configuration.

\section*{QUANTIFICATION AND STATISTICAL ANALYSIS}

Sensitivity analysis (one-at-a-time) was carried out to see the effects of parameters in the numerical model.

\end{document}